\documentclass[twocolumn]{aastex61}
\pdfoutput=1 
\usepackage{amsmath,amstext}
\usepackage[T1]{fontenc}
\usepackage{apjfonts} 
\usepackage[figure,figure*]{hypcap}
\usepackage{color}

\usepackage{booktabs,threeparttable} 
\usepackage{verbatim} 
\usepackage{rotating} 
\usepackage{url} 

\shorttitle{Multiscale fragmentation in the Perseus molecular cloud}
\shortauthors{R. Pokhrel et al.}

\begin{document}

\title{Hierarchical fragmentation in the Perseus molecular cloud: From the cloud scale to protostellar objects}

\author{Riwaj Pokhrel}
\affiliation{Department of Astronomy, University of Massachusetts, Amherst, MA 01003, USA}
\affiliation{Harvard-Smithsonian Center for Astrophysics, 60 Garden Street, Cambridge, MA 02138, USA}
\author{Philip C. Myers}
\affiliation{Harvard-Smithsonian Center for Astrophysics, 60 Garden Street, Cambridge, MA 02138, USA}
\author{Michael M. Dunham}
\affiliation{Department of Physics, State University of New York at Fredonia, 280 Central Ave, Fredonia, NY 14063, USA}
\affiliation{Harvard-Smithsonian Center for Astrophysics, 60 Garden Street, Cambridge, MA 02138, USA}
\author{Ian W. Stephens}
\affiliation{Harvard-Smithsonian Center for Astrophysics, 60 Garden Street, Cambridge, MA 02138, USA}
\author{Sarah I. Sadavoy}
\affiliation{Harvard-Smithsonian Center for Astrophysics, 60 Garden Street, Cambridge, MA 02138, USA}
\author{Qizhou Zhang}
\affiliation{Harvard-Smithsonian Center for Astrophysics, 60 Garden Street, Cambridge, MA 02138, USA}
\author{Tyler L. Bourke}
\affiliation{Harvard-Smithsonian Center for Astrophysics, 60 Garden Street, Cambridge, MA 02138, USA}
\affiliation{SKA Organization, Jodrell Bank Observatory, Lower Withington, Macclesfield, Cheshire SK11 9DL, UK}
\author{John J. Tobin}
\affiliation{Leiden Observatory, Leiden University, P.O. Box 9513, 2300-RA Leiden, The Netherlands}
\affiliation{Department of Physics and Astronomy, University of oklahoma, 440 W. Brooks St., Norman, OK 73019, USA}
\author{Katherine I. Lee}
\affiliation{Harvard-Smithsonian Center for Astrophysics, 60 Garden Street, Cambridge, MA 02138, USA}
\author{Robert A. Gutermuth}
\affiliation{Department of Astronomy, University of Massachusetts, Amherst, MA 01003, USA}
\author{Stella S. R. Offner}
\affiliation{Department of Astronomy, University of Texas at Austin, Austin, TX 78712, USA}

\begin{abstract}
We present a study of hierarchical structure in the Perseus molecular cloud, from the scale of the entire cloud ($\gtrsim$10 pc) to smaller clumps ($\sim$1 pc), cores ($\sim$0.05-0.1 pc), envelopes ($\sim$300-3000 AU) and protostellar objects ($\sim$15 AU). We use new observations from the Submillimeter Array (SMA) large project "Mass Assembly of Stellar Systems and their Evolution with the SMA (MASSES)" to probe the envelopes, and recent single-dish and interferometric observations from the literature for the remaining scales. This is the first study to analyze hierarchical structure over five scales in the same cloud complex. We compare the number of fragments with the number of Jeans masses in each scale to calculate the Jeans efficiency, or the ratio of observed to expected number of fragments. The velocity dispersion is assumed to arise either from purely thermal motions, or from combined thermal and non-thermal motions inferred from observed spectral line widths. For each scale, thermal Jeans fragmentation predicts more fragments than observed, corresponding to inefficient thermal Jeans fragmentation. For the smallest scale, thermal plus non-thermal Jeans fragmentation also predicts too many protostellar objects. However at each of the larger scales thermal plus non-thermal Jeans fragmentation predicts fewer than one fragment, corresponding to no fragmentation into envelopes, cores, and clumps. Over all scales, the results are inconsistent with complete Jeans fragmentation based on either thermal or thermal plus non-thermal motions. They are more nearly consistent with inefficient thermal Jeans fragmentation, where the thermal Jeans efficiency increases from the largest to the smallest scale.
\end{abstract}


\section{Introduction} \label{intro}

Fragmentation in molecular clouds has been well studied over many years \citep{Larson78,Miyama84,Monaghan91,Rodriguez05,Contreras16,Li17}. Fragmentation is a process that produces "fragments" or structures in a molecular cloud. A hierarchy of nested structures is often created by the process of hierarchical fragmentation as seen in some recent observation and simulation studies (see \citealt{Dobbs14} and \citealt{Heyer15} for recent reviews). These studies show that clouds, which are typically $\gtrsim$10 pc in size have a wide range of structures from larger filaments and clumps to dense cores and disks.

\begin{figure}[h]
\centering
\includegraphics[scale=0.3]{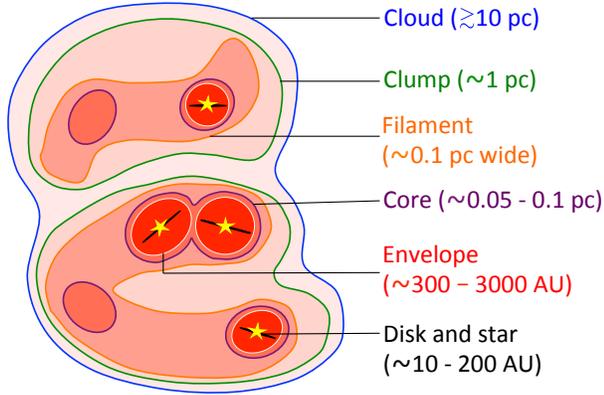}
\caption{A cartoon display of a molecular cloud showing hierarchical structures inside the cloud. The figure shows the cloud, clumps, filaments, cores, envelopes, and protostellar systems that we consider in this study. The image is not drawn to scale.
\label{hierarchy}
}
\end{figure}

Figure \ref{hierarchy} summarizes the scales and terms we utilize for this analysis in a cartoon of the hierarchical structures in a molecular cloud. We use  "cloud" to identify the largest structure of our interest on scales of $\gtrsim$10 pc. A cloud fragments into "clumps" which are $\sim$1 pc in size \citep{Ridge06, Sadavoy14}. Inside the clumps, we observe elongated gaseous filaments that are $\sim$0.1 pc wide \citep{Arzoumanian11}. Inside the filaments we find $\sim$0.05-0.1 pc cores \citep{di07} which are the sites where new stars are able to form. In this paper we report the detection of further dense condensations of size scale $\sim$300-3000 AU which we term "envelopes". Dense, inner envelopes or protostellar disks surrounding a central young star are often found inside the envelope. Disks have a range of size from $<$10 AU (B335; \citealt{Yen15}) to $>$200 AU (L1448IRS3B; \citealt{Looney00,Tobin16}).

\begin{figure*}
\centering
\includegraphics[scale=0.61]{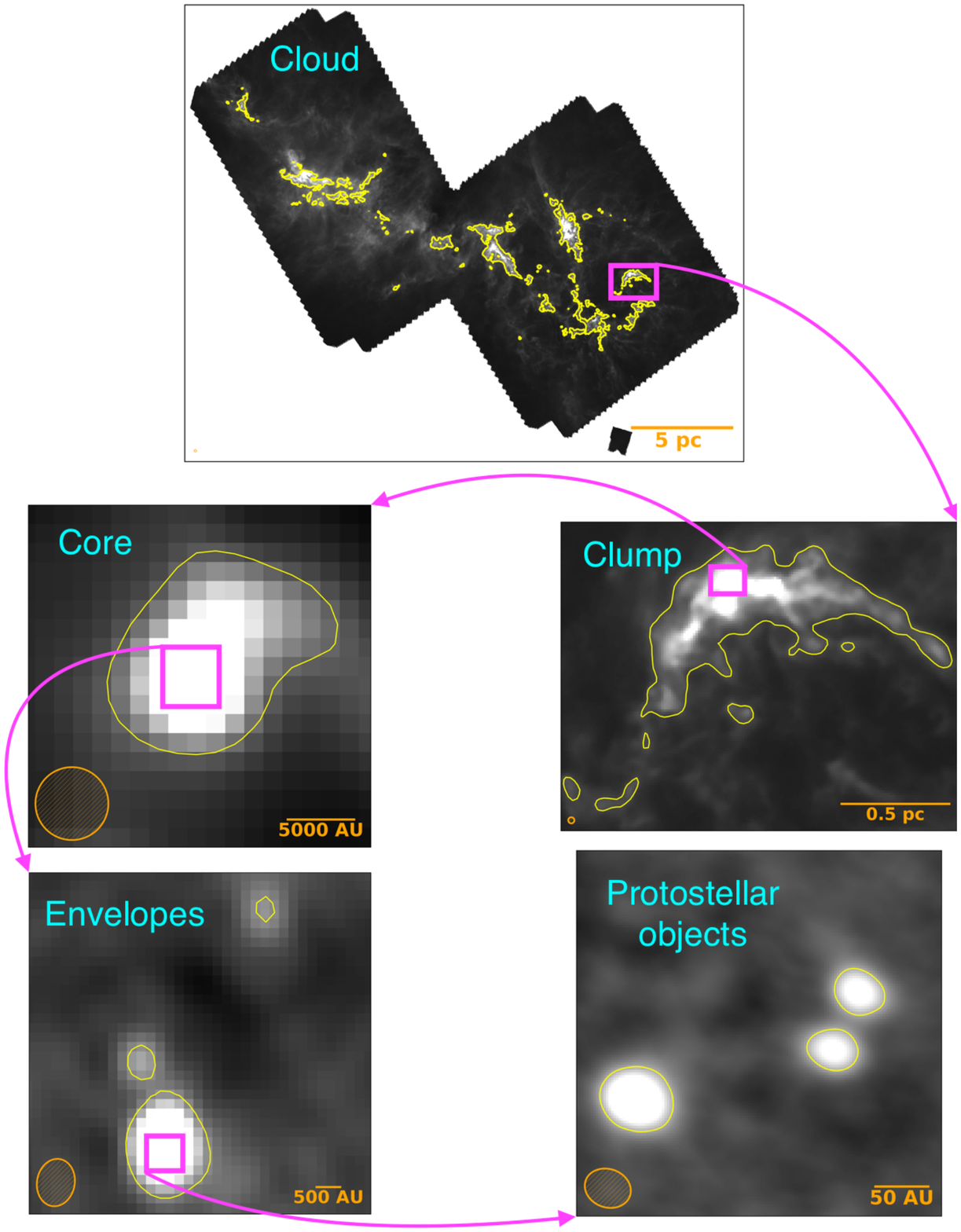}
\caption{Multi-scale structures in the Perseus molecular cloud. In each panel, beam size is shown in lower left and scale is shown in lower right. The five different panels are explained below.\\
\textbf{ \underline{Cloud:}} The entire Perseus cloud at 350 $\micron$ obtained from \emph{Herschel}. Yellow contours correspond to A$_V$ = 7 mag (see \citealt{Sadavoy14}) and are derived from the opacity map from \cite{Zari16}. The coordinates of the center of map are $R.A.$(J2000) = 3h35m06.08s \& $Dec$(J2000) = +31d24m10.61s. The FWHM beam size is 24.9$''$.\\
\textbf{ \underline{Clump:}} One of the clumps from Herschel 350 $\micron$ map, L1448 is magnified to show the details. Yellow contour shows A$_V$ = 7 mag (see Panel Cloud). The coordinates of the center of map are $R.A.$(J2000) = 3h25m25.91s \& $Dec$(J2000) = +30d38m47.91s. The FWHM beam size is 24.9$''$.\\
\textbf{ \underline{Core:}} SCUBA 850 $\micron$ map of one of the cores (J032536.1+304514) that resides in L1448 (map from \citealt{Di08}). Yellow contour represents a 5$\sigma$ level where $\sigma$ = 0.1 Jy/beam. The coordinates of the center of map are $R.A.$(J2000) = 3h25m35.77s \& $Dec$(J2000) = +30d45m25.49s. The FWHM beam size is $\sim$23$''$.\\
\textbf{ \underline{Envelopes:}} SMA 1.3 mm map of the region that is shown by magenta box in Panel Core. The yellow contours represent 6$\sigma$ detection, where $\sigma$ = 0.012 Jy/beam. The coordinates of the center of map are $R.A.$(J2000) = 3h25m31.15s \&1 $Dec$(J2000) = +30d45m23.89s. The angular resolution of this map is  $\sim$ 4$''$ $\times$ 3$''$.\\
\textbf{ \underline{Protostellar object:}} VLA map from VANDAM survey \citep{Tobin16} for one of the envelope `Per-emb-33'. Yellow contours represent 15$\sigma$ limit (see \citealt{Lee15}) where $\sigma$ = 7.25 $\mu$Jy/beam. The coordinates of the center of map are $R.A.$(J2000) = 3h25m36.34s \& $Dec$(J2000) = +30d45m15.07s. The FWHM beam size is 0.065$''$.
\label{hierarchy_perseus}
}
\end{figure*}

Figure \ref{hierarchy_perseus} displays the hierarchical structures in the Perseus molecular cloud from actual observations. The figure includes 5 panels where each panel represents structures of varying size scales starting from the largest structure in our study, the whole cloud, and moving subsequently towards smaller structures such as clumps, cores, envelopes and protostellar objects. The first panel "Cloud" shows the larger scale $Herschel$ 350 $\micron$ emission map where 7 clumps are detected (see \citealt{Sadavoy14,Mercimek17}). In one of the clumps, L1448 \citep{Terebey97, Looney00, Kwon06}, \cite{Sadavoy10} found the presence of four cores (three protostellar and one starless) from SCUBA observations \citep{Di08}. One of the cores, J032536.1+304514 inside L1448 when observed with the SMA revealed three envelope scale fragments. Observations from the VLA show the presence of three protostellar objects in one of the SMA detected envelopes, Per-emb-33 \citep{Lee15, Tobin16}.

The multi-scale structures in a molecular cloud can be produced by a variety of fragmentation processes. Some of these processes include magnetohydrodynamic turbulence (e.g., \citealt{MacLow04}, \citealt{Hennebelle12}), self-gravity of the gas (e.g., \citealt{Heyer09,Ballesteros11,Ballesteros12}) and ionization radiation (e.g., \citealt{Whitworth94},\citealt{Dale09}). According to the turbulence regulated star formation theory \citep{Padoan99,MacLow04,Krumholz05}, supersonic turbulence creates a series of density fluctuations, where long-lasting high density fluctuations are able to gravitationally collapse. In self-gravity regulated star formation theory, cloud fragmentation is dominated by gravity, and gravity rather than turbulence is responsible for the structure hierarchy (e.g., \citealt{Hoyle53}, \citealt{Zinnecker84}, \citealt{Heitsch08}, \citealt{Ballesteros11}). In some cases, the colliding clouds produce initial turbulence which creates non-uniform density distribution, and then gravity takes over (gravo-turbulent fragmentation; \citealt{Klessen04}). Although what controls the fragmentation process is still debated, it is likely some combination of gravitational instability, turbulence, magnetic fields, and stellar feedback (e.g., \citealt{Padoan02}, \citealt{Hosking04}, \citealt{Machida05}, \citealt{Girart13}).

In terms of support, gas thermal pressure is expected to be the most important factor against gravitational collapse on the smaller scales relevant to the formation of individual stars \citep{Larson06}. At these scales, cloud fragmentation is expected to follow classical Jeans instability that is obtained by balancing gravity with thermal pressure \citep{Jeans29}. If the actual mass of a cloud is greater than its Jeans mass, self gravity wins over the thermal support and the cloud fragments. Another prevailing view is that self-gravitating clouds are supported against collapse by non-thermal motions \citep{Heitsch00,Clark05} rather than the thermal support. For this case, the non-thermal motions provides the pressure necessary to balance the inward pull of gravity.

This study stands out when compared to other similar studies regarding cloud fragmentation for mainly two reasons. First, we focus on hierarchical fragmentation over multiple scales in the same cloud, rather than combining observations from various different clouds. Thus we have a uniform sampling region and same physical conditions at each scale. Second, this study covers the entire cloud down to the scale of protostellar objects. Previous analyses were unable to probe well these small scales because of limitations in observational techniques. Hence, this is the first study to investigate a detailed hierarchical fragmentation picture in a single molecular cloud from the scale of the cloud to the scale of protostellar objects.

We explain our observations in \S \ref{obs} where we describe our new SMA observations as well as the complementary data from the literature. In \S \ref{result} we present the newly identified SMA sources. In \S \ref{jeansanalysis}, we present the Jeans analysis for each level of hierarchy. In \S \ref{combination}, we combine all the hierarchies for a comprehensive study. We discuss our results in \S \ref{conclusions} and finally we present our conclusions in \S \ref{summary}.

\subsection{Target selection}

The Perseus molecular cloud ($d$ = 230 pc, \citealt{Hirota08,Hirota11}) is an ideal target for this analysis. It is one of the best studied nearby star forming regions with ample data available in the literature, including observations at mid-IR (\emph{Spitzer}), far-IR (\emph{Herschel}) and sub-mm (JCMT, CSO) wavelengths. These observations probe the warm dust emission from young stars as well as cooler dust from the ambient cloud and its dense clumps and cores. The Perseus protostars have also been probed with the VLA \cite{Tobin16} at the scales of protostellar disks. Finally, Perseus has a relatively large population of young stars compared to other nearby molecular clouds.  Since we want to focus on the hierarchical structure, it is advantageous to examine younger populations that are still embedded in their natal environment.  Thus, Perseus provides a large, unbiased  sample necessary to obtain the statistics for this study.

\section{Observations} \label{obs}

\subsection{Archival Data} \label{archivaldata}

Our study spans spatial scales from $\gtrsim$15 AU to $\gtrsim$10 pc. To observe this multiscale structure, we require data from multiple telescopes including both single dish telescopes and interferometers. 

For the cloud scales, we used global properties of Perseus from near-infrared extinction maps in \cite{Sadavoy10}. For clump scales, we used the physical properties determined in \cite{Sadavoy14} from observations with the {\emph Herschel Space Observatory} \citep{Pilbratt10} at far-IR wavelengths.

For core scales, we used the source lists provided in \cite{Sadavoy10} and \cite{Mercimek17} at submillimeter wavelengths from the Submillimeter Common-User Bolometer Array (SCUBA; \citealt{Holland99}) at the James Clerk Maxwell Telescope. The cores were initially identified from the SCUBA Legacy Catalogue \citep{Di08} and classified as starless or protostellar using infrared observations from Spitzer (see \citealt{Sadavoy10} for details).

Finally, for disk scales, we used the results from the "VLA Nascent Disk and Multiplicity" survey (VANDAM; PI: J. Tobin) undertaken with the Karl G. Jansky Very Large Array (VLA; \citealt{Thompson80}) at 8 mm \citep{Tobin16}. These data probed all protostars in Perseus at a common, high resolution of 15 AU. At this spatial resolution, VANDAM sources probe the dense gas and dust immediately surrounding the protostars. The VLA sources represent scales from protostellar vicinity to compact dust disks. For the purpose of this study, we term all such VLA sources as "protostellar objects". Thus by "protostellar objects" we encompass the size scales from protostars to compact disks.

As noted above, we have literature data for the scale of the entire cloud, clumps, cores and disks for the Perseus molecular cloud. However, we lack the data for the envelope scales. The MASSES data from the SMA (see \S \ref{smaobservation}) fill that gap and enables us to study envelope scale structures.

\subsection{SMA observations} \label{smaobservation}

\subsubsection{MASSES}

We used observations from the large-scale SMA project ($\sim$600 observing hours, 3-4 years) "Mass Assembly of Stellar Systems and Their Evolution with the SMA" (MASSES; co-PIs: M. Dunham and I. Stephens). MASSES targeted all known 73 protostars in Perseus in dust continuum and spectral line emission at 230 and 345 GHz. The data were taken in the sub-compact (SUB) and extended (EXT) array configurations. The SUB configuration has an angular resolution of $\sim$ 4$''$ at 230 GHz, which corresponds to a spatial scale of $\sim$1000 AU at the distance of Perseus. The EXT configuration has an angular resolution of $\sim$ 1$''$ at 230 GHz ($\sim$200 AU). The MASSES observations include line emission at $^{12}$CO (2-1), $^{13}$CO (2-1), C$^{18}$O (2-1) \& N$_2$D$^+$ (3-2) at 230 GHz. We do not include the line data in this study. We also do not discuss the 345 GHz (0.87 mm) data at this time and instead focus on the 230 GHz (1.3 mm) results.


The VANDAM and MASSES projects target the same protostars in Perseus and complement each other. Nevertheless, the MASSES data at 1.3 mm are better able to trace the envelope emission than the VANDAM data at 8 mm, because thermal dust emission is brighter at 1.3 mm than 8 mm by two orders of magnitude. Due to this limitation, the VANDAM data will primarily trace material associated with the very inner envelope and disk \citep{Tobin16} where the densities are highest rather than the surrounding envelope. Thus, the SMA data presented here are key to trace the envelope scales of our analysis.

For this study, we used only 230 GHz continuum data observed in the SUB configuration. The data were observed with the ASIC correlator with 2 GHz bandwidth in each of the lower and upper sidebands. Each 2 GHz band has 24 chunks with 82 MHz usable bandwidth. Our correlator setup includes 8 chunks with 64 channels in each chunk for continuum observations. The remaining chunks are used for line observations. We averaged the chunks with 64 channels per chunk to generate the continuum. The continuum thus generated has an effective bandwidth of 1312 MHz considering both the upper and lower sidebands. 


\subsubsection{SMA Data Reduction}

We used the MIR software package\footnote{https://www.cfa.harvard.edu/~cqi/mircook.html} with standard calibration procedures to reduce and calibrate the visibility data. First, we did the baseline correction on the visibility dataset and flagged the bad data points. We then corrected the amplitude and phase data with the system temperature. We calibrated bandpass using antenna based solutions for the bandpass calibrator which is then followed by the gain calibration and ultimately the flux calibration using bright quasars or planets. Typically we used the quasar 3c84 for gain calibration, either 3c84, 1058+015, 3c454.3 or a similar bright quasar for bandpass calibration, and Uranus for flux calibration. The uncertainty in the flux calibration is $\sim$25\% (see \citealt{Lee15}).

We used the MIRIAD software package \citep{Sault95} to image the calibrated visibility data. After taking the inverse Fourier transform of the visibility data, the image was obtained using the robust parameter = 1 with MIRIAD task $clean$. This provided the midway solution of both the natural and uniform weighting, enabling the detection of both small scale structures and extended emission. The images were cleaned and restored until finally they were corrected for primary beam attenuation using an image of the primary beam pattern.

\section{SMA Results} \label{result}

\subsection{SMA source identification} \label{smasource}

\begin{figure}[tbh]
\centering
\includegraphics[scale=0.3, angle=270]{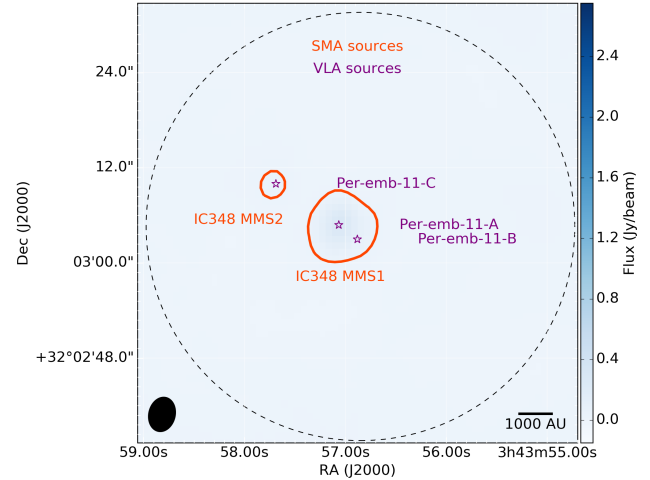}
\caption{The VLA detected sources (protostellar objects shown by purple stars) are overplotted in the SMA image (SMA envelopes are shown by 5$\sigma$ orange contours) in the case of Per-emb-11. The two SMA sources are IC348 MMS1 and IC348 MMS2, and the three VLA sources are Per-emb-11-A, Per-emb-11-B and Per-emb-11-C. The angular resolution size is shown at lower left and scale bar is shown at lower right respectively. Dash circle represents primary beam of the pointing.
\label{per11}
}
\end{figure}

For the purpose of this study, we defined an SMA source $(envelope)$ as a source that is detected at $>$ 5$\sigma$, where $\sigma$ is the noise in the background image. Figure \ref{per11} shows an example of SMA sources at 5$\sigma$ level that are detected in the region of Per-emb-11. We overplotted the higher resolution VLA sources in the reduced SMA tracks, which are shown as purple stars. The figure shows two SMA sources, "IC348 MMS1" and "IC348 MMS2". The first source "IC348 MMS1" contains two VLA sources, Per-emb-11-A and Per-emb-11-B. The second source "IC348 MMS2" contains only one VLA source, Per-emb-11-C. The nomenclatures IC348 MMS1 and MMS2  for SMA sources are adopted from \cite{Lee16}. Images corresponding to all the SMA-detected sources will be publicly available in FITS (Flexible Image Transport System) format in the online version of this paper.

We found a total of 73 SMA sources in the Perseus molecular cloud. To avoid duplications of the same source from different tracks, we excluded the detections that are far from the center of primary beam. After excluding the duplicated sources, we had a total of 56 unique SMA sources (53 sources at $>$ 5$\sigma$ and 3 sources at $>$ 6$\sigma$ level). We list these sources in Table \ref{envelopes}. There are also 3 unique detections at $>$ 4$\sigma$ give in Table \ref{envelopes} which we consider robust enough detections for further analysis. Thus, we identify 59 distinct sources with the SMA in the Perseus molecular cloud.

\subsection{SMA Source fitting} \label{sourcefitting}

We calculated SMA source sizes by fitting models of each source in the visibility plane. The reason we chose to fit in visibility plane instead of the image plane is because some of the SMA sources had extended structure. These structures are better seen in visibilities and in some instances are not adequately recovered after we inverse Fourier Transformed the visibility data and deconvolved the dirty image from the dirty beam. For example, we found that source sizes were generally underestimated when fit in the image plane over the visibility plane because of spatial filtering. Thus, we calculated the source sizes in the visibility plane.
 
\begin{figure}[ht]
\centering
\vskip -1.0in
\includegraphics[scale=0.4, angle=0]{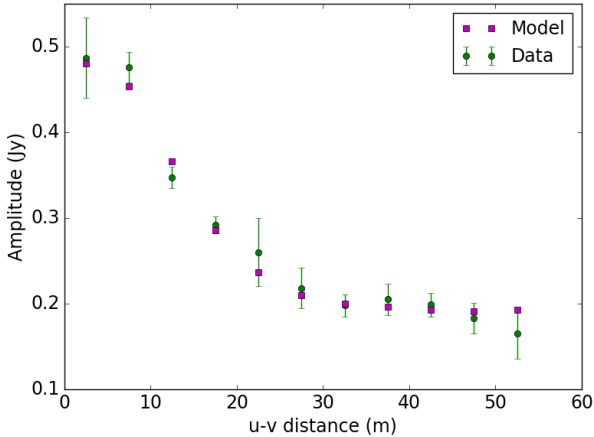}
\vskip -1.0in
\caption{Radial profiles of amplitude with u-v distance for IC348 MMS1. The green circles represent actual visibility data with 3-$\sigma$ error bars on noise (before taking flux calibration error into account). The data is fit by a model that is a combination of a Gaussian function and a point function. This model is shown by magenta squares. The position of the source is determined by fitting the source in the image plane before fitting them in visibility plane.
\label{ampvsuvdist}
}
\end{figure}

To determine the best fit model that describes the nature of the source, we inspected plots of the amplitude with $uv$ distance ($amp$ versus $uvdist$). If the variation of amplitude with u-v distance showed a Gaussian nature, we fitted a Gaussian model to the source since the Fourier transform of a Gaussian function is also a Gaussian function (but of a varying width). Similarly if visibility amplitude is constant across the range of $uv$ distance, we fitted the source with a point source as the Fourier transform of a uniform function is a point source. Finally if the variation showed a Gaussian nature with a uniform tail, we fitted a combined model of a point and a Gaussian function. Figure \ref{ampvsuvdist} shows an example of a combined fit in the case of IC348 MMS1 (one of the two SMA detected sources in Per-emb-11 in Figure \ref{per11}).

In the cases of multiple sources in the same field, we need to specify the location and flux of each source separately in the visibility plane. To estimate such source properties, first we used the MIRIAD routine $imfit$ to find source position and flux in the image plane. Then we used them as initial guesses while using MIRIAD task $uvfit$ to fit the sources in the visibility plane. Our technique of source fitting works in MIRIAD as long as there are less than 20 initial free parameters because of restrictions in \emph{uvfit}. If there are more than 20 initial free parameters, we reduced the number of sources by subtracting a source in the image plane and again obtained the fits for the residual $u-v$ data in the visibility plane. In brief, first we transformed the actual visibility data to the image plane. Then we cleaned the data and restored the clean map by deconvolving with the dirty beam. We identified the source that we want to subtract. After subtracting the source, we Fourier transformed the residual image data back to the visibility plane and fitted the remaining continuum sources. We repeated the process by subtracting other sources to cross check the consistency in values of fitted parameters. We plotted the best fit models on top of the continuum images and visually confirmed that these were indeed good fits.

\startlongtable
\begin{deluxetable*}{ccccccccc}
\centering
\tabletypesize{\scriptsize}
\tablecolumns{9}
\tablewidth{0pc}
\tablecaption{SMA source properties obtained by fitting the source \label{envelopes}}
\tablehead{\colhead{SMA source} & \colhead{Fitting model} & \colhead{R.A.\tablenotemark{(b)}} & \colhead{Dec.\tablenotemark{(b)}} & \colhead{Peak flux\tablenotemark{(c)}} & \colhead{Integrated flux\tablenotemark{(c)}} & \colhead{Major axis\tablenotemark{(d)}} & \colhead{Minor axis\tablenotemark{(d)}} & \colhead{Group\tablenotemark{(e)}}\\
\colhead{name\tablenotemark{(a)}} & \colhead{} & \colhead{(J2000)} & \colhead{(J2000)} & \colhead{(mJy)} & \colhead{(mJy)} & \colhead{($''$)} & \colhead{($''$)} & \colhead{}
}
\startdata
B1-bN & Point + Gaussian & 03:33:21.198 & +31:07:43.931 & 152.7 $\pm$ 3.7 & 248.7 $\pm$ 11.4 & 6.41 $\pm$ 1.24 & 6.04 $\pm$ 1.53 & A \\
IC348 MMS1\tablenotemark{(f)} & Point + Gaussian & 03:43:57.055 & +32:03:04.669 & 195.6 $\pm$ 2.6 & 477.9 $\pm$ 7.6 & 6.71 $\pm$ 0.22 & 5.56 $\pm$ 0.22 & A \\
IC348 MMS2\tablenotemark{(f)} & Point + Gaussian & 03:43:57.735 & +32:03:10.098 & 23.1 $\pm$ 3.4 & 78.8 $\pm$ 6.7 & 5.41 $\pm$ 0.72 & 3.13 $\pm$ 0.77 & B \\
IRAS4B$'$ & Point + Gaussian & 03:29:12.825 & +31:13:06.962 & 227.1 $\pm$ 166.0 & 311.6 $\pm$ 232.6 & 1.97 $\pm$ 2.6 & 0.51 $\pm$ 2.6 & B \\
L1448IRS3\tablenotemark{(g)} & Point + Gaussian & 03:25:35.675 & +30:45:35.163 & 51.9 $\pm$ 2.7 & 337.4 $\pm$ 12.2 & 12.96 $\pm$ 0.48 & 4.77 $\pm$ 0.23 & A \\
L1448NW\tablenotemark{(g)} & Point + Gaussian & 03:25:36.464 & +30:45:21.425 & 105.1 $\pm$ 2.8 & 218.8 $\pm$ 9.5 & 10.87 $\pm$ 1.0 & 3.48 $\pm$ 1.0 & A \\
L1451-MMS & Point & 03:25:10.241 & +30:23:55.013 & 39.1 $\pm$ 3.1 & 39.1 $\pm$ 3.1 & ... & ... & B \\
Per-bolo-45-SMM\tablenotemark{(h)} & Point + Gaussian & 03:29:06.764 & +31:17:22.297 & 7.5 $\pm$ 3.9 & 108.6 $\pm$ 24.1 & 14.21 $\pm$ 2.39 & 9.44 $\pm$ 2.39 & A \\
Per-bolo-58 & Gaussian & 03:29:25.417 & +31:28:14.205 & 24.3 $\pm$ 5.2 & 94.5 $\pm$ 24.4 & 14.27 $\pm$ 1.41 & 7.54 $\pm$ 0.99 & A \\
Per-emb-1 & Point + Gaussian & 03:43:56.770 & +32:00:49.865 & 118.7 $\pm$ 2.7 & 331.6 $\pm$ 6.9 & 6.65 $\pm$ 0.25 & 4.64 $\pm$ 0.25 & A \\
Per-emb-2 & Point + Gaussian & 03:32:17.915 & +30:49:48.033 & 350.6 $\pm$ 9.0 & 764.7 $\pm$ 12.0 & 3.54 $\pm$ 0.12 & 2.89 $\pm$ 0.08 & B \\
Per-emb-3 & Point & 03:29:00.554 & +31:11:59.849 & 59.5 $\pm$ 3.0 & 59.5 $\pm$ 3.0 & ... & ... & B \\
Per-emb-5 & Point + Gaussian & 03:31:20.931 & +30:45:30.334 & 206.3 $\pm$ 3.5 & 329.0 $\pm$ 6.4 & 5.98 $\pm$ 0.39 & 3.85 $\pm$ 0.39 & A \\
Per-emb-8 & Point + Gaussian & 03:44:43.975 & +32:01:34.968 & 111.2 $\pm$ 2.7 & 183.1 $\pm$ 10.7 & 8.91 $\pm$ 1.88 & 7.71 $\pm$ 1.88 & A \\
Per-emb-9 & Point + Gaussian & 03:29:51.876 & +31:39:05.516 & 15.8 $\pm$ 3.4 & 174.2 $\pm$ 25.0 & 13.3 $\pm$ 1.47 & 10.76 $\pm$ 1.47 & A \\
Per-emb-10 & Point + Gaussian & 03:33:16.412 & +31:06:52.384 & 13.6 $\pm$ 1.9 & 58.8 $\pm$ 8.9 & 9.42 $\pm$ 1.57 & 7.58 $\pm$ 1.57 & A \\
Per-emb-10-SMM & Point + Gaussian & 03:33:18.470 & +31:06:33.629 & 4.9 $\pm$ 1.8 & 19.1 $\pm$ 4.0 & 4.01 $\pm$ 2.43 & 3.99 $\pm$ 2.43 & B \\
Per-emb-12 & Point + Gaussian & 03:29:10.490 & +31:13:31.369 & 1484.0 $\pm$ 14.8 & 4093.0 $\pm$ 22.8 & 4.4 $\pm$ 0.05 & 3.28 $\pm$ 0.04 & B \\
Per-emb-13 & Point + Gaussian & 03:29:11.993 & +31:13:08.137 & 687.0 $\pm$ 13.6 & 1173.5 $\pm$ 18.1 & 4.1 $\pm$ 0.18 & 3.24 $\pm$ 0.15 & B \\
Per-emb-14 & Point + Gaussian & 03:29:13.517 & +31:13:57.754 & 87.0 $\pm$ 4.6 & 123.3 $\pm$ 7.9 & 4.24 $\pm$ 1.35 & 1.22 $\pm$ 1.35 & B \\
Per-emb-15 & Point + Gaussian & 03:29:04.207 & +31:14:48.642 & 8.2 $\pm$ 4.0 & 69.2 $\pm$ 12.1 & 8.34 $\pm$ 1.6 & 5.24 $\pm$ 1.36 & A \\
Per-emb-16 & Point + Gaussian & 03:43:50.999 & +32:03:23.858 & 11.0 $\pm$ 2.3 & 93.6 $\pm$ 10.8 & 9.12 $\pm$ 1.17 & 7.73 $\pm$ 1.17 & A \\
Per-emb-17 & Point + Gaussian & 03:27:39.120 & +30:13:02.526 & 47.7 $\pm$ 3.0 & 116.5 $\pm$ 11.4 & 9.65 $\pm$ 1.6 & 6.24 $\pm$ 1.59 & A \\
Per-emb-18 & Point + Gaussian & 03:29:11.261 & +31:18:31.326 & 117.4 $\pm$ 4.0 & 217.9 $\pm$ 16.6 & 8.71 $\pm$ 1.57 & 7.67 $\pm$ 1.38 & A \\
Per-emb-19 & Point & 03:29:23.476 & +31:33:28.940 & 14.7 $\pm$ 2.5 & 14.7 $\pm$ 2.5 & ... & ... & B \\
Per-emb-19-SMM\tablenotemark{(h)} & Point & 03:29:24.331 & +31:33:22.569 & 8.9 $\pm$ 2.6 & 8.9 $\pm$ 2.6 & ... & ... & B \\
Per-emb-20 & Gaussian & 03:27:43.199 & +30:12:28.962 & 1.1 $\pm$ 1.0 & 53.8 $\pm$ 16.7 & 9.56 $\pm$ 1.25 & 3.93 $\pm$ 0.91 & A \\
Per-emb-20-SMM & Gaussian & 03:27:42.778 & +30:12:25.936 & 7.2 $\pm$ 0.9 & 14.8 $\pm$ 16.4 & 3.59 $\pm$ 1.31 & 0.02 $\pm$ 84.0 & B \\
Per-emb-21 & Point + Gaussian & 03:29:10.688 & +31:18:20.151 & 43.6 $\pm$ 4.1 & 193.6 $\pm$ 14.3 & 7.14 $\pm$ 1.37 & 6.41 $\pm$ 1.28 & A \\
Per-emb-22 & Point + Gaussian & 03:25:22.353 & +30:45:13.213 & 92.8 $\pm$ 3.9 & 400.4 $\pm$ 13.0 & 8.28 $\pm$ 0.48 & 6.2 $\pm$ 0.47 & A \\
Per-emb-23 & Point + Gaussian & 03:29:17.249 & +31:27:46.336 & 12.4 $\pm$ 1.9 & 78.5 $\pm$ 8.8 & 12.57 $\pm$ 1.49 & 7.2 $\pm$ 1.04 & A \\
Per-emb-25 & Point & 03:26:37.492 & +30:15:27.904 & 87.8 $\pm$ 3.7 & 87.8 $\pm$ 3.7 & ... & ... & B \\
Per-emb-26 & Point + Gaussian & 03:25:38.872 & +30:44:05.299 & 180.1 $\pm$ 2.3 & 480.6 $\pm$ 13.0 & 11.62 $\pm$ 0.46 & 7.28 $\pm$ 0.25 & A \\
Per-emb-27 & Point + Gaussian & 03:28:55.562 & +31:14:37.167 & 259.6 $\pm$ 2.8 & 709.7 $\pm$ 7.8 & 6.85 $\pm$ 0.15 & 5.61 $\pm$ 0.14 & A \\
Per-emb-28 & Point + Gaussian & 03:43:50.987 & +32:03:07.967 & 12.0 $\pm$ 2.0 & 58.8 $\pm$ 12.0 & 11.89 $\pm$ 2.89 & 8.12 $\pm$ 2.2 & A \\
Per-emb-29 & Point + Gaussian & 03:33:17.860 & +31:09:32.307 & 144.2 $\pm$ 3.6 & 468.2 $\pm$ 11.7 & 7.88 $\pm$ 0.3 & 5.98 $\pm$ 0.26 & A \\
Per-emb-30 & Point & 03:33:27.302 & +31:07:10.187 & 50.9 $\pm$ 3.9 & 50.9 $\pm$ 3.9 & ... & ... & B \\
Per-emb-33\tablenotemark{(g)} & Point + Gaussian & 03:25:36.324 & +30:45:14.771 & 495.1 $\pm$ 5.8 & 1050.7 $\pm$ 8.8 & 5.1 $\pm$ 0.13 & 3.47 $\pm$ 0.13 & A \\
Per-emb-35 & Point + Gaussian & 03:28:37.124 & +31:13:31.236 & 43.6 $\pm$ 2.9 & 127.2 $\pm$ 10.4 & 9.7 $\pm$ 2.13 & 6.14 $\pm$ 1.65 & A \\
Per-emb-36 & Point + Gaussian & 03:28:57.363 & +31:14:15.610 & 129.3 $\pm$ 1.9 & 220.8 $\pm$ 11.5 & 13.69 $\pm$ 1.45 & 6.83 $\pm$ 0.82 & A \\
Per-emb-37 & Point + Gaussian & 03:29:18.936 & +31:23:13.109 & 12.0 $\pm$ 2.0 & 59.1 $\pm$ 7.3 & 10.63 $\pm$ 1.45 & 5.5 $\pm$ 1.21 & A \\
Per-emb-40 & Point & 03:33:16.646 & +31:07:54.808 & 25.3 $\pm$ 13.5 & 25.3 $\pm$ 13.5 & ... & ... & B \\
Per-emb-41 & Point + Gaussian & 03:33:21.338 & +31:07:26.439 & 285.5 $\pm$ 4.1 & 374.6 $\pm$ 11.4 & 5.86 $\pm$ 1.65 & 5.56 $\pm$ 1.65 & A \\
Per-emb-44 & Point + Gaussian & 03:29:03.719 & +31:16:03.295 & 333.4 $\pm$ 3.2 & 759.1 $\pm$ 17.7 & 9.16 $\pm$ 0.44 & 4.56 $\pm$ 0.22 & A \\
Per-emb-47 & Point & 03:28:34.513 & +31:00:50.702 & 9.2 $\pm$ 2.4 & 9.2 $\pm$ 2.4 & ... & ... & B \\
Per-emb-50 & Point & 03:29:07.764 & +31:21:57.162 & 96.4 $\pm$ 2.9 & 96.4 $\pm$ 2.9 & ... & ... & B \\
Per-emb-51 & Point + Gaussian & 03:28:34.521 & +31:07:05.467 & 12.1 $\pm$ 5.4 & 115.4 $\pm$ 10.7 & 5.77 $\pm$ 0.84 & 3.69 $\pm$ 0.71 & A \\
Per-emb-53 & Point + Gaussian & 03:47:41.577 & +32:51:43.745 & 24.9 $\pm$ 4.0 & 74.3 $\pm$ 9.9 & 6.9 $\pm$ 1.67 & 5.02 $\pm$ 1.26 & A \\
Per-emb-54 & Point + Gaussian & 03:29:02.828 & +31:20:41.321 & 21.7 $\pm$ 4.6 & 197.4 $\pm$ 13.0 & 10.48 $\pm$ 0.68 & 5.9 $\pm$ 0.52 & A \\
Per-emb-56 & Point & 03:47:05.422 & +32:43:08.330 & 14.1 $\pm$ 6.1 & 14.1 $\pm$ 6.1 & ... & ... & B \\
Per-emb-57 & Point & 03:29:03.322 & +31:23:14.338 & 23.3 $\pm$ 1.4 & 23.3 $\pm$ 1.4 & ... & ... & B \\
Per-emb-58\tablenotemark{(h)} & Point & 03:28:58.361 & +31:22:16.811 & 7.7 $\pm$ 1.6 & 7.7 $\pm$ 1.6 & ... & ... & B \\
Per-emb-61 & Point & 03:44:21.301 & +31:59:32.526 & 11.6 $\pm$ 3.9 & 11.6 $\pm$ 3.9 & ... & ... & B \\
Per-emb-62 & Point & 03:44:12.973 & +32:01:35.289 & 75.8 $\pm$ 3.1 & 75.8 $\pm$ 3.1 & ... & ... & B \\
Per-emb-63 & Point & 03:28:43.279 & +31:17:33.248 & 18.2 $\pm$ 2.9 & 18.2 $\pm$ 2.9 & ... & ... & B \\
Per-emb-64 & Point & 03:33:12.848 & +31:21:23.950 & 45.7 $\pm$ 24.3 & 45.7 $\pm$ 24.3 & ... & ... & B \\
Per-emb-65 & Point & 03:28:56.301 & +31:22:27.693 & 27.5 $\pm$ 2.6 & 27.5 $\pm$ 2.6 & ... & ... & B \\
SVS13B & Point + Gaussian & 03:29:03.032 & +31:15:51.362 & 248.6 $\pm$ 3.2 & 774.5 $\pm$ 19.3 & 8.97 $\pm$ 0.44 & 6.75 $\pm$ 0.18 & A \\
SVS13C & Point + Gaussian & 03:29:01.969 & +31:15:38.199 & 55.1 $\pm$ 3.1 & 189.0 $\pm$ 11.1 & 12.12 $\pm$ 0.97 & 3.49 $\pm$ 0.97 & A \\
\enddata
\tablenotetext{(a)}{ The SMA source names are adopted from \cite{Tobin16} for consistency with previous nomenclature. For some of the Per-emb sources, we detected a secondary source with the SMA that could not be found in literature. For these sources, we added the suffix "SMM" to the end of the name. For example, Per-bolo-45-SMM, does not lie in the same region as Per-bolo-45. All the SMA sources are detected at 5-$\sigma$ contour, unless otherwise stated.}
\tablenotetext{(b)}{ R.A. and Dec. refers to the peak position of SMA source obtained by fitting a model to the source (see \S \ref{sourcefitting}).}
\tablenotetext{(c)}{ The reported uncertainties are statistical and they exclude any calibration/systematic error.}
\tablenotetext{(d)}{ Deconvolved FWHM size estimates with the model synthesized beam.}
\tablenotetext{(e)}{ Group "A": Size estimates in both image and visibility plane agree, axes size /  axes error $>$ 3. Group "B": Either one or both of these conditions are not met.}
\tablenotetext{(f)}{ Nomenclature adopted from \cite{Lee16}.}
\tablenotetext{(g)}{ Source is detected at 6$\sigma$ contour.}
\tablenotetext{(h)}{ Source is detected at 4$\sigma$ contour.}
\end{deluxetable*}

Not all SMA sources are robust even if they are detected at $>$ 5$\sigma$. For example, the sources that we fit with only a point function are unresolved point sources and thus do not have size estimates, we use the resolution limit as an upper limit on size. Other sources are not well fit by the models and have large uncertainties in their axis ratios. Based on these possible sources of errors, in Table \ref{envelopes} we divided the SMA sources into 2 groups, "A" where the fitting results are trustworthy and can be considered for further analyses, and "B" where the fitting results may have systematic errors and are not robust. There are 34 SMA sources that belong to group "A" and 25 SMA sources belong to group "B". For the sources that belong to group "A", the sizes estimated in both the image plane and the visibility plane are within 10 percent of each other. For our main analyses, we focus on the group "A" sources.

To calculate the peak and integrated flux of an SMA source, we fitted the same model (that we obtain for that source in the visibility plane) in the primary beam corrected SMA map. These flux estimates are used to determine the masses of the SMA sources in \S \ref{mass estimation}.

\subsection{SMA versus VLA multiplicity} \label{sourceproperties}

Figure \ref{per11} shows an example where multiplicity is seen at the scale for both SMA envelopes and VLA protostellar objects. The observed multiplicity at different scales raises an important question of whether or not the multiplicity seen at the larger scales in the previous generation (envelopes) are transferred to the smaller scales in next generation (disk scale and protostellar objects). To study this, we have counted the multiplicity for both SMA envelopes and VLA protostellar objects for all the available samples. 

The number counting of SMA and VLA sources are defined by the resolution limit and the primary beam of the observation. Hence the SMA sources are counted within 1,000 AU and 10,000 AU and the VLA sources are counted within 15 AU and 1,000 AU. For the purpose of counting sources, each SMA field is centered at the center of the primary beam (c.f., Figure \ref{per11}). We consider only those SMA and VLA sources that lie within the primary beam of the SMA image to have a consistency in the number of sources. For the sources that lie in more than one primary beam (overlapping beams), we only include the source what is close to the center of the primary beam and discard the ones that are away from the center of primary beam, as those regions are prone to be less sensitive and noisier. This way we do not end up counting the same source more than once and have a consistent sample of sources.

\begin{figure}[tbh]
\centering
\vskip -1.0in
\includegraphics[scale=0.4, angle=0]{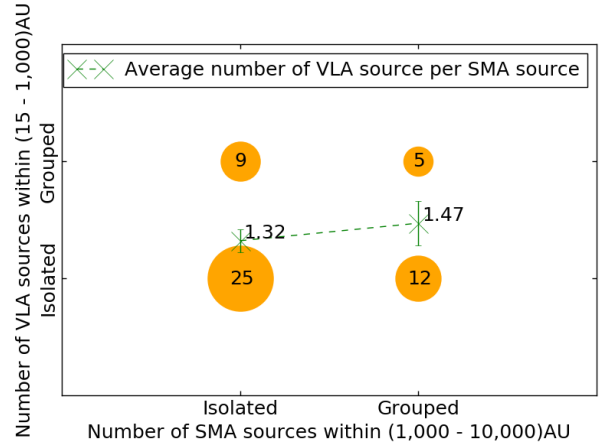}
\vskip -1.0in
\caption{X-axis shows the number of SMA sources between 1,000 AU and 10,000 AU that are either single (isolated) or multiple (grouped), and y-axis shows the number of VLA sources between 10 AU and 1,000 AU that are either isolated or grouped. The SMA sources are detected with at least 5$\sigma$ contour. The scale in each axis is determined by the resolution limit and the primary beam of the respective telescope array. The sizes of the yellow circles are proportional to the number of SMA envelopes (written inside yellow markers). The dash green line connects the average number of VLA sources per SMA source. 
\label{smavdm}
}
\end{figure}

The multiplicity at scales of both the SMA and VLA sources are shown in Figure \ref{smavdm}. In Figure \ref{smavdm}, we have differentiated the SMA and VLA sources into four categories. The first category contains the isolated SMA source that has an isolated VLA source inside. We had 25 such cases. The second category includes isolated SMA sources that have multiple or grouped ($>$1) VLA sources. We had 9 such sources. The third category contains SMA sources that are grouped within 1,000-10,000 AU but single isolated VLA source in them. We had 12 such cases. The fourth category contains grouped SMA sources that have multiple VLA sources. We had 5 such cases. For an isolated SMA source, there are an average of 1.32 VLA sources, and for the grouped SMA sources there are an average of 1.47 VLA sources (shown by green cross in Figure \ref{smavdm}). The isolated and grouped SMA objects show relatively equal numbers of VLA objects (within errors), although there are hints it could be increasing. Hence, the trend in Figure \ref{smavdm} is limited by statistical uncertainty.

\section{Multi-scale Jeans Analysis} \label{jeansanalysis}

As discussed in \S \ref{intro}, the most accepted means of external support to a cloud structure against the gravitational pull is the thermal support, the turbulent support, and the support due to magnetic fields.
For the foregoing SMA and VLA sources, and for the larger regions which enclose them, we tested the observed hierarchical structures under two possible Jeans fragmentation cases. First, we assume that the structures are supported entirely by thermal gas motions. Next, we assume that the structure is supported by the combined effect of both thermal and non-thermal motions. These two cases are useful because they may be considered simple lower and upper limits to the true level of support against gravitational fragmentation and collapse. The non-thermal motions adopted here from observed line widths are simpler than those in numerical simulations of MHD turbulent fragmentation, which are more anisotropic, time-varying, and scale-dependent (\citealt{Padoan99,Padoan02,Hennebelle11,hopkins13}). Although the terms  "non-thermal" and "turbulent" are often used interchangeably, to avoid confusion in this paper we refer to the motions inferred from line widths as "non-thermal" motions. Our non-thermal Jeans analysis simply tests whether turbulence can act as an isotropic pressure, rather than testing turbulence fragmentation models.

A gas cloud is said to be Jeans stable against fragmentation when the outward thermal pressure exerted by gas motion balances the inward gravitational pull of the cloud. If the inward gravitational force wins over the outward thermal balance, the system becomes Jeans unstable and can fragment. The critical mass when the cloud becomes unstable is called the Jeans mass (M$_{\rm{J}}$). We used Equation \ref{eq:1} to calculate the Jeans mass assuming a spherical geometry at all the levels of the cloud hierarchy and also assuming that the Jeans length represents the diameter of the sphere \citep{Binney87}, i,e.,
\begin{equation} \label{eq:1}
M_{\rm{J}} = \frac{\pi^{5/2}}{6G^{3/2}}c_{\rm{eff}}^3\rho_{\rm{eff}}^{-1/2},
\end{equation}
where $c_{\rm{eff}}$ is the `effective sound speed', $G$ is the universal gravitational constant, and $\rho_{\rm{eff}}$ is the average density of the region assuming spherical geometry.

For the first case of a pure thermal support to a cloud structure, thermal Jeans mass $M_{\rm{J}}^{\rm{th}}$ is calculated assuming $c_{\rm{eff}}$ same as the thermal sound speed, $c_{\rm{s}}$, which is calculated as,
\begin{equation}\label{soundspeed}
c_{\rm{s}} = \sqrt{\frac{\gamma k_{\rm{B}} T}{\mu_{H_2} m_{\rm{H}}}},
\end{equation}
where $\gamma$ is the adiabatic constant which is unity for an isothermal medium, $k_{\rm{B}}$ is the Boltzmann constant, $T$ is the average temperature of the region, $\mu_{H_2}$ is the mean molecular weight per hydrogen molecule ($\sim$2.8 for a cloud with 71\% molecular hydrogen, 27\% helium and 2\% metals; \citealt{Kauffmann08}), $m_{\rm{H}}$ is hydrogen mass.

For the second case, we applied an upper limit to the thermal fragmentation.  Here, we adopted ``thermal temperatures'' based on the combined support from both thermal and non-thermal gas motions. We used different molecular line tracers from the literature to trace gas motions at all scales.  For each tracer, we used the observed velocity dispersion of the line ($\sigma^{\rm{obs}}$), which is comprised of both thermal ($\sigma^{\rm{th}}$) and non-thermal ($\sigma^{\rm{nth}}$) components. We then calculated the non-thermal component of the lines by subtracting out the thermal velocity dispersion, i.e., $\sigma^{\rm{nth}}$ = $\sqrt{(\sigma^{\rm{obs}})^2 - (\sigma^{\rm{th}})^2}$, using $\sqrt{k_{\rm{B}} T/\mu m_{\rm{H}}}$ for the thermal velocity dispersion and the appropriate molecular weight, $\mu$, for each tracer (for example 29 for $^{13}$CO, 17 for NH$_3$). Finally, we added, in quadrature, the non-thermal line widths to the thermal sound speed, i.e., $\sigma^{\rm{th,nth}}$ = $\sqrt{c_{\rm{s}}^2 + (\sigma^{\rm{nth}})^2}$, and used this combined velocity dispersion ($\sigma^{\rm{th,nth}}$) to calculate the Jeans mass in Equation \ref{eq:5}. For the system that is supported by both thermal and non-thermal motions, the Jeans mass is given as (see \citealt{Palau14,Palau15}),

\begin{equation}\label{eq:5}
\Bigg[\frac{M_{\rm{J}}^{\rm{th,nth}}}{M_{\odot}} \Bigg] = 0.8 \Bigg[ \frac{\sigma^{\rm{th,nth}}}{0.19~\rm{kms^{-1}}} \Bigg]^3 \Bigg[\frac{n_{\rm{H_2}}}{10^5~\rm{cm^{-3}}} \Bigg]^{-1/2}
\end{equation}

For both conditions of support, the expected number of fragments that are produced in a structure in any generation is given by the ratio of total mass of the structure to the Jeans mass of the same structure. This ratio is also called the Jeans number and is calculated as

\begin{equation}
N_{\rm{J}} = \frac{M_{\rm{total}}}{M_{\rm{J}}}.
\end{equation}


We have studied the possibility of Jeans fragmentation for the observed multi-scale substructures in the Perseus molecular cloud. We performed this analysis in a hierarchical fashion from the cloud scale to the scale of protostellar objects in Perseus (the approximate size-scale of each structure is shown in Figure \ref{hierarchy}). The fragmenting scale is hereafter called the parent structure and its subsequent fragments are hereafter child structures. For example, if cloud is the parent structure then clump is the child structure, and so on. 

We define the formation efficiency of fragments as the ratio of the number of child or child structures to the Jeans number of the parent structure. This definition is similar to the core formation efficiency (CFE), which is defined as the ratio of the number of cores detected in a clump to the Jeans number of that particular clump (\citealt{Bontemps10, Palau15}). Since the children are formed from the available mass of the parent structure, the formation efficiency of a child structure can not be greater than one.

\subsection{Cloud to Clump} \label{cloudtoclump}

For the Perseus molecular cloud, the largest scale fragmentation is the cloud to clump scale. Perseus has a mass of 3.3 $\times$ 10$^4$ M$_{\odot}$ and covers an area of roughly 66 deg$^2$ above extinction $A_V$ = 1 \citep{Sadavoy10}. These measurements assume a different distance to the cloud and hence for consistency the measurements are corrected for 230 pc distance. The cloud has been studied extensively in dust and molecular line emission to identify its clumps \citep{Ridge06,Sadavoy14,Zari16}. Clumps are relatively dense parsec scale structures that are often defined as the regions in which most stars form (regions within $A_V$ $\sim$ 7 mag, \citealt{Andre10,Lada10,Evans14}). Based on this definition, there are seven clumps in the Perseus cloud \citep{Sadavoy14,Mercimek17}.

For our Jeans analysis of the Perseus cloud, we first assumed that only thermal pressure is supporting the cloud against its self-gravitation. \cite{Zari16} gives a line-of-sight average temperature map for the Perseus cloud from modified blackbody fits to thermal dust emission. Based on this temperature map, we adopted the average dust temperature of 18 K to use in the our Jeans analysis. The transition between atomic and molecular form takes place between $A_V$ $\sim$1 and 2, so we perform the Jeans analysis in cloud where $A_V$ > 2. The corresponding density for $A_V$ = 2 in Perseus molecular cloud is 200 cm$^{-3}$ \citep{Evans09}. Using these parameters, we get thermal Jeans mass $\sim$35 M$_{\odot}$ for the Perseus cloud. The corresponding mass at Av $>$ 2 is $\sim$4000 M$_{\odot}$ which gives thermal Jeans number $\sim$120 using the cloud mass above. This Jeans number far exceeds the observed number of clumps (7), and leads to a clump formation efficiency in Perseus of only 0.06. 

Molecular clouds, however, are unlikely to be supported against fragmentation by solely thermal pressure.  In particular, clouds show substantial non-thermal motions that can provide additional support.  For example, $^{13}$CO observations in Perseus \citep{Ridge03, Kirk10} have a typical velocity dispersion of 0.9 kms$^{-1}$ whereas the thermal line width of this molecule is expected to be $<$ 0.1 kms$^{-1}$. The non-thermal motions are predominantly present at the cloud scale as inferred from the typical velocity dispersion of 0.9 kms$^{-1}$ from \cite{Kirk10}. The total Jeans mass using $\sigma^{\rm{th,nth}}$ = 0.9 kms$^{-1}$ is $\sim$2000 M$_{\odot}$, assuming a typical cloud density of 200 cm$^{-3}$ for material at Av $>$ 2, which is appropriate for tracing $^{13}$CO \citep{Evans09}. Similarly, we find a Jeans number of 2 and a Jeans efficiency of 3.8. This efficiency greater than unity is not physical. There are additional factors like magnetic fields which can provide support in the low density environment of clouds that have not been considered in this analysis.

\subsection{Clump to Core} \label{clumptocore}

For the second level of hierarchy, we explored the scale from clumps to cores. Cores of size scale $\sim$0.1 pc reside in the clumps. \cite{Sadavoy10} used SCUBA (850 $\micron$) and \emph{Spitzer Space Telescope} (3.6-70 $\micron$) to explore the dense cores in Perseus. They classified the sub-mm cores that were found with SCUBA as starless or protostellar using point source photometry from $Spitzer$ wide field surveys (see \citealt{Sadavoy10} for details). The details of individual starless and protostellar cores in each clump are presented in \cite{Sadavoy10}. \cite{Mercimek17} characterized the distribution of these cores inside the clumps. 

Similar to the previous hierarchy, first we tested the expected number of thermal Jeans fragments against the observed number of fragments. To calculate the Jeans number of the clumps, we used the line-of-sight averaged temperatures and mass derived in \cite{Sadavoy14}. 
Table \ref{clumps} gives the Jeans masses, numbers, and efficiencies for each clump assuming pure thermal support. We use the mass and areas from \cite{Mercimek17} to determine the average density of each clump for $A_{\rm{V}}$ $>$ 7 mag and the dust temperatures from \cite{Sadavoy14} to estimate the thermal support. 

The velocity dispersion at the scales where $A_{\rm{V}}$ $>$ 7 mag can be studied by using C$^{18}$O line width. The typical line width in Perseus from C$^{18}$O is 0.4 kms$^{-1}$ \citep{Hatchell05}. We used this average velocity dispersion to find $\sigma^{\rm{th,nth}}$ and estimate the Jeans parameters assuming that both thermal and non-thermal motions are supporting the stability of clumps. 

Table \ref{clumps} also gives an estimate of the Jeans mass, Jeans number and Jeans efficiency for each clump assuming this combined thermal and non-thermal case. We find the values of $\epsilon^{\rm{th}}$ between 0.06 and 0.6, similar to the independent estimates of CFE by \cite{Palau15} using a different sample of objects and observations. We find an average $\epsilon^{\rm{th}}$ of 0.2. For the combined support, the CFE is $>$ 1 for most of the clumps.

\begin{figure}[tbh]
\centering
\vskip -1.0in
\includegraphics[scale=0.4]{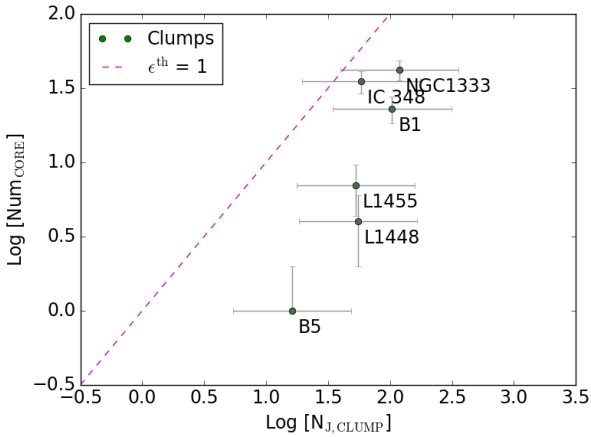}
\vskip -1.0in
\caption{Comparison between number of enclosed cores with the Jeans number of the clumps. The error in number of cores assume Poisson statistics and the Jeans number is correct within a factor of 3. 
\label{NjVsN_cl}
}
\end{figure}

Figure \ref{NjVsN_cl} compares the number of enclosed cores in each clump ($Num_{\rm{CORE}}$) with the corresponding Jeans number of the clumps ($N_{\rm{J, CLUMP}}$). The plot shows that the number of cores increases with the Jeans number of the clumps (Pearson's correlation coefficient = 0.8). This agreement suggests that thermal Jeans fragmentation may play a significant role in forming cores. Nevertheless, there are systematically fewer cores than predicted, which suggests that thermal pressure is not sufficient.

In Figure \ref{NjVsN_cl}, we consider Poisson statistics in estimating the uncertainty in the number of cores. Thus the uncertainty in the number of cores is given by the square root of that number, which is an upper limit of uncertainty. For the Jeans number of the clumps, the sources of uncertainty are mass, temperature and area of the clump. However, uncertainty in mass is the dominant source of error (correct within a factor of a few). We propagated uncertainty on the dependent variables and found that the Jeans number is uncertain up to a factor of 3, if we a take factor of 2 as the lower limit mass uncertainty. This is true for all other levels of hierarchy as well so we have implemented the same technique for error estimates in other hierarchies.

\begin{deluxetable*}{lccccccccc}
\centering
\tablecolumns{10}
\tablecaption{Jeans analysis in the clumps \label{clumps}}
\tablehead{\colhead{Clump} & \colhead{Mass\tablenotemark{(a)}} & \colhead{Area\tablenotemark{(a)}} & \colhead{M$_{\rm{J}}$$^{\rm{th}}$} & \colhead{M$_{\rm{J}}$$^{\rm{th,nth}}$} & \colhead{N$_{\rm{J}}$$^{\rm{th}}$} & \colhead{N$_{\rm{J}}$$^{\rm{th,nth}}$} &  \colhead{Num$_{\rm{CORE}}$} & \colhead{$\epsilon^{\rm{th}}$\tablenotemark{(b)}} & \colhead{$\epsilon^{\rm{th,nth}}\tablenotemark{(b)}$}\\
\colhead{} & \colhead{[M$_{\odot}$]} & \colhead{[pc$^2$]} & \colhead{[M$_{\odot}$]} & \colhead{[M$_{\odot}$]} & \colhead{} & \colhead{} & \colhead{} & \colhead{} & \colhead{}}
\startdata
B5 & 62 & 0.32 & 3.8 & 41.7 & 16.2 & 1.5 & 1 & 0.06 & 0.67 \\
B1-E & 88 & 0.57 & 5.1 & 54.3 & 17.2 & 1.6 & 0 & 0.0 & 0.0 \\
L1448 & 159 & 0.48 & 2.9 & 34.6 & 55.1 & 4.6 & 4 & 0.07 & 0.87 \\
L1455 & 251 & 1.3 & 4.7 & 57.9 & 53.1 & 4.3 & 7 & 0.13 & 1.61 \\
IC 348 & 511 & 2.9 & 8.7 & 79.3 & 58.6 & 6.4 & 35 & 0.6 & 5.43 \\
NGC1333 & 568 & 2.0 & 4.8 & 54.0 & 119.0 & 10.5 & 42 & 0.35 & 4.0 \\
B1 & 598 & 2.5 & 5.8 & 62.8 & 103.9 & 9.5 & 23 & 0.22 & 2.41 \\
\enddata
\tablenotetext{(a)}{ For the regions that are contoured by an equivalent of A$_{v}$ $>$ 7 mag in $Herschel$ derived column density maps \citep{Mercimek17}.}
\tablenotetext{(b)}{ Efficiency is calculated by taking ratio of the number of cores to the Jeans number of clumps considering both thermal ($\epsilon^{\rm{th}}$) and combined ($\epsilon^{\rm{th,nth}}$) support.}
\end{deluxetable*}

\subsection{Core to envelope} \label{coretoenvelope}

At a next level of hierarchy, we explored the scales of cores to envelopes (see Figures \ref{hierarchy} and \ref{hierarchy_perseus} for the difference between core and envelope scales). The properties of cores are discussed in \S \ref{clumptocore}. For the envelopes, we used the SMA observations from MASSES discussed in \S \ref{sourcefitting} and \ref{sourceproperties}.

\begin{figure}[tbh]
\centering
\includegraphics[scale=0.32, angle = 270]{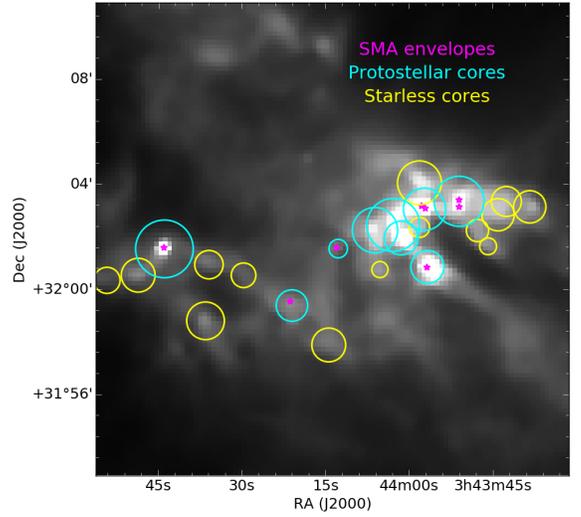}
\caption{Positions of cores and envelopes in IC 348. The cyan circles are protostellar cores, yellow circles are starless cores and magenta stars are the SMA envelopes. Background image is 350$\micron$ $Herschel$ dust emission map.
\label{envMapping}
}
\end{figure}

To estimate the number of envelopes present in each core, we examined the spatial correspondence between the SMA envelopes and the SCUBA cores. Figure \ref{envMapping} shows the distribution of cores and envelopes in IC 348. The mass and area of cores are taken from \cite{Sadavoy10}. The positions of SMA envelopes are the peak positions obtained by fitting the sources as explained in \S \ref{sourcefitting}. To determine whether or not the envelopes are spatially coincident with the dense cores, we used a set of boundary conditions as outlined below.

First, we found the core that is closest to the given envelope. Second, we used a minimum distance criterion to identify whether or not the envelope is associated with its nearest core. For simplicity, we consider an envelope associated with a core if it is within one core radius of the core center, where radius is taken to be same as the effective radius. This effective radius is calculated from the area of core by assuming a spherical geometry ($\sqrt[]{A/\pi}$). Applying the selection criteria, we found either 0, 1, 2 or 3 envelopes inside a single core by counting the number of SMA sources. If an envelope is expected in a core from pre-existing data \citep{Enoch09} but is not detected with the SMA, we consider that core to have 0 envelopes for consistency.

\begin{figure}[tbh]
\centering
\vskip -1.0in
\includegraphics[scale=0.4]{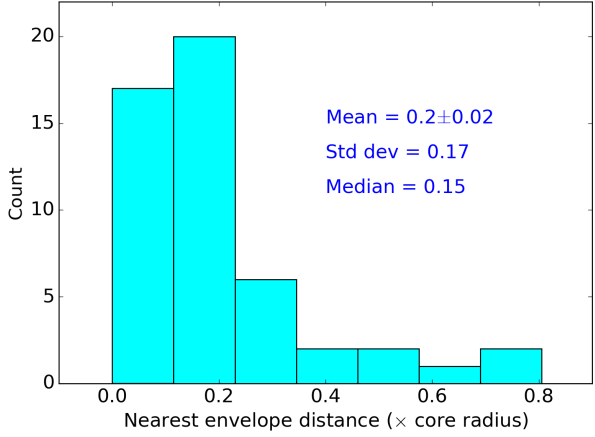}
\vskip -1.0in
\caption{Distribution of the nearest envelope distance between the envelopes and the cores in terms of the core radii.
\label{nnd}
}
\end{figure}

The minimum envelope distance is calculated in terms of core radii by dividing the distance between the centers of SMA envelope and its nearest core by the radius of that core. Figure \ref{nnd} represents the histogram of the minimum envelope distance. The mean and median of the histogram is $\sim$0.2 and 0.15, showing that the envelopes lie mostly around the center of core. This degree of central concentration is highly significant compared to a random distribution of envelopes within cores. This is consistent with \cite{Jorgensen07} where they find that young stars are primarily found in the interiors of dense cores.

\cite{Rosolowsky08} measured the velocity dispersion in cores and core candidates in the Perseus molecular cloud using the ammonia observations with Green Bank Telescope (GBT). They find a typical gas kinetic temperature of $\sim$11 K and a median velocity dispersion of $\sim$0.18 kms$^{-1}$. We used these values and the core properties from \cite{Sadavoy10} to perform Jeans analysis for the core and envelope scales.

Table \ref{cores} summarizes the Jeans instability in the Perseus cores for both thermal support and combined thermal and non-thermal support. The Table lists only the cores where envelopes were sampled by the SMA so it doesn't represent all the cores in Perseus (see \citealt{Sadavoy10} for all the SCUBA detected cores in Perseus). The average envelope formation efficiency for a thermally supported core ($\epsilon^{\rm{th}}$) is $\sim$0.4, and for the combined support ($\epsilon^{\rm{th,nth}}$) it is $\sim$1.

\begin{deluxetable*}{lccccccccc}
\centering
\tabletypesize{\scriptsize}
\tablecolumns{10}
\tablecaption{Jeans analysis in the cores \label{cores}}
\tablehead{\colhead{Core} & \colhead{Mass\tablenotemark{(a)}} & \colhead{Area\tablenotemark{(a)}} & \colhead{M$_{\rm{J}}$$^{\rm{th}}$} & \colhead{M$_{\rm{J}}$$^{\rm{th,nth}}$} & \colhead{N$_{\rm{J}}$$^{\rm{th}}$} & \colhead{N$_{\rm{J}}$$^{\rm{th,nth}}$} &  \colhead{Num$_{\rm{ENVELOPE}}$} & \colhead{$\epsilon^{\rm{th}}$\tablenotemark{(b)}} & \colhead{$\epsilon^{\rm{th,nth}}\tablenotemark{(b)}$}\\
\colhead{} & \colhead{[M$_{\odot}$]} & \colhead{[pc$^2$]} & \colhead{[M$_{\odot}$]} & \colhead{[M$_{\odot}$]} & \colhead{} & \colhead{} & \colhead{} & \colhead{} & \colhead{}}
\startdata
J032522.2+304514 & 3.6 & 0.00478 & 0.5 & 1.2 & 7.4 & 3.0 & 1 & 0.14 & 0.34 \\
J032536.1+304514 & 17.3 & 0.00985 & 0.4 & 1.0 & 44.1 & 17.8 & 3 & 0.07 & 0.17 \\
J032538.9+304402 & 4.9 & 0.0043 & 0.4 & 1.0 & 12.5 & 5.0 & 1 & 0.08 & 0.2 \\
J032739.2+301259 & 2.0 & 0.00283 & 0.5 & 1.1 & 4.3 & 1.7 & 1 & 0.23 & 0.57 \\
J032742.9+301228 & 2.1 & 0.00385 & 0.6 & 1.4 & 3.8 & 1.5 & 2 & 0.53 & 1.31 \\
J032832.2+311108 & 2.0 & 0.00694 & 0.9 & 2.2 & 2.3 & 0.9 & 0 & 0.0 & 0.0 \\
J032834.5+310702 & 0.4 & 0.00102 & 0.5 & 1.2 & 0.7 & 0.3 & 1 & 1.38 & 3.42 \\
J032836.9+311326 & 3.7 & 0.00724 & 0.7 & 1.7 & 5.4 & 2.2 & 1 & 0.18 & 0.46 \\
J032839.2+310556 & 3.1 & 0.00785 & 0.8 & 1.9 & 4.0 & 1.6 & 0 & 0.0 & 0.0 \\
J032845.2+310549 & 1.4 & 0.00454 & 0.8 & 1.9 & 1.7 & 0.7 & 0 & 0.0 & 0.0 \\
J032855.2+311437 & 12.4 & 0.01208 & 0.5 & 1.3 & 23.0 & 9.3 & 2 & 0.09 & 0.22 \\
J032900.3+311201 & 0.8 & 0.00212 & 0.6 & 1.5 & 1.3 & 0.5 & 1 & 0.79 & 1.96 \\
J032901.3+312031 & 15.1 & 0.01094 & 0.5 & 1.1 & 33.3 & 13.4 & 1 & 0.03 & 0.07 \\
J032903.6+311455 & 7.5 & 0.00817 & 0.5 & 1.3 & 14.5 & 5.8 & 1 & 0.07 & 0.17 \\
J032906.9+311725 & 1.5 & 0.00229 & 0.4 & 1.1 & 3.5 & 1.4 & 1 & 0.28 & 0.71 \\
J032907.4+312155 & 5.9 & 0.00817 & 0.6 & 1.4 & 10.1 & 4.1 & 1 & 0.1 & 0.24 \\
J032910.1+311331 & 24.5 & 0.00636 & 0.2 & 0.6 & 103.6 & 41.7 & 1 & 0.01 & 0.02 \\
J032910.7+311824 & 10.8 & 0.01169 & 0.6 & 1.4 & 19.2 & 7.7 & 2 & 0.1 & 0.26 \\
J032912.0+311306 & 15.2 & 0.00754 & 0.3 & 0.8 & 44.5 & 17.9 & 2 & 0.04 & 0.11 \\
J032913.4+311354 & 4.6 & 0.00528 & 0.5 & 1.2 & 9.7 & 3.9 & 1 & 0.1 & 0.26 \\
J032917.4+312748 & 3.3 & 0.0095 & 0.9 & 2.2 & 3.8 & 1.5 & 1 & 0.27 & 0.66 \\
J032918.7+312312 & 1.4 & 0.00264 & 0.5 & 1.3 & 2.7 & 1.1 & 1 & 0.37 & 0.92 \\
J032925.4+312818 & 0.9 & 0.00246 & 0.6 & 1.5 & 1.6 & 0.6 & 1 & 0.63 & 1.56 \\
J032951.4+313904 & 1.4 & 0.00342 & 0.6 & 1.5 & 2.3 & 0.9 & 1 & 0.44 & 1.1 \\
J033120.7+304531 & 3.4 & 0.00608 & 0.6 & 1.5 & 5.4 & 2.2 & 1 & 0.18 & 0.46 \\
J033217.6+304947 & 7.2 & 0.0095 & 0.6 & 1.5 & 12.3 & 4.9 & 1 & 0.08 & 0.2 \\
J033313.2+311956 & 2.1 & 0.00581 & 0.7 & 1.9 & 2.9 & 1.2 & 0 & 0.0 & 0.0 \\
J033315.9+310656 & 14.6 & 0.01496 & 0.6 & 1.4 & 25.1 & 10.1 & 1 & 0.04 & 0.1 \\
J033316.4+310750 & 6.2 & 0.00916 & 0.6 & 1.5 & 9.9 & 4.0 & 1 & 0.1 & 0.25 \\
J033317.8+310932 & 17.8 & 0.02488 & 0.8 & 1.9 & 23.0 & 9.3 & 1 & 0.04 & 0.11 \\
J033318.2+310608 & 1.4 & 0.00478 & 0.8 & 2.0 & 1.7 & 0.7 & 1 & 0.6 & 1.49 \\
J033321.0+310732 & 17.5 & 0.01327 & 0.5 & 1.2 & 36.0 & 14.5 & 2 & 0.06 & 0.14 \\
J033327.1+310707 & 3.0 & 0.00785 & 0.8 & 2.0 & 3.8 & 1.5 & 1 & 0.26 & 0.65 \\
J034351.0+320321 & 6.1 & 0.01057 & 0.7 & 1.7 & 8.7 & 3.5 & 2 & 0.23 & 0.57 \\
J034356.7+320051 & 10.0 & 0.01094 & 0.6 & 1.4 & 18.0 & 7.3 & 1 & 0.06 & 0.14 \\
J034357.2+320303 & 6.9 & 0.00694 & 0.5 & 1.2 & 14.5 & 5.8 & 2 & 0.14 & 0.34 \\
J034401.4+320157 & 3.4 & 0.00554 & 0.6 & 1.4 & 5.9 & 2.4 & 0 & 0.0 & 0.0 \\
J034412.7+320133 & 0.1 & 0.00021 & 0.4 & 1.0 & 0.1 & 0.1 & 1 & 7.92 & 19.65 \\
J034421.0+315923 & 2.3 & 0.00882 & 1.0 & 2.5 & 2.3 & 0.9 & 1 & 0.44 & 1.08 \\
J034443.9+320132 & 3.5 & 0.00754 & 0.7 & 1.8 & 4.9 & 2.0 & 1 & 0.2 & 0.5 \\
\enddata
\tablenotetext{(a)}{ \cite{Sadavoy10}.}
\tablenotetext{(b)}{ Efficiency is calculated by taking ratio of the number of envelopes to the Jeans number of cores considering both thermal ($\epsilon^{\rm{th}}$) and combined ($\epsilon^{\rm{th,nth}}$) support.}
\end{deluxetable*}

\begin{figure}[tbh]
\centering
\vskip -1.0in
\includegraphics[scale=0.4]{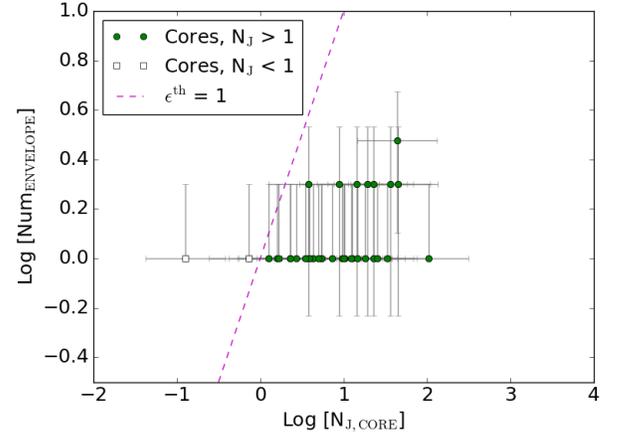}
\vskip -1.0in
\caption{Comparison of the number of enclosed envelopes with the Jeans number of the parent cores considering pure thermal Jeans analysis. The green circles have Jeans number > 1 and the hollow squares have Jeans number < 1. The magenta dash line represents $\epsilon^{\rm{th}}$ = 1 relation. The uncertainty in the number of enclosed envelopes follow Poisson statistics, which is an upper limit uncertainty. Jeans number of cores are uncertain within a factor of 3.
\label{NjVsN_core}
}
\end{figure}

Figure \ref{NjVsN_core} shows the number of enclosed envelopes with the thermal Jeans number of their parent cores with the same format as in Figure \ref{NjVsN_cl} for cores in clumps. The magenta dash line represents $\epsilon^{\rm{th}} = 1$ line where thermal Jeans fragmentation predicts the exact number of fragments. The relation between the number of enclosed envelopes and the Jeans number of the cores is hard to constrain because of the high uncertainties. Nevertheless, the average number of envelopes is less than that predicted by the thermal Jeans analysis of the cores.

\begin{figure}[tbh]
\centering
\includegraphics[scale=0.42]{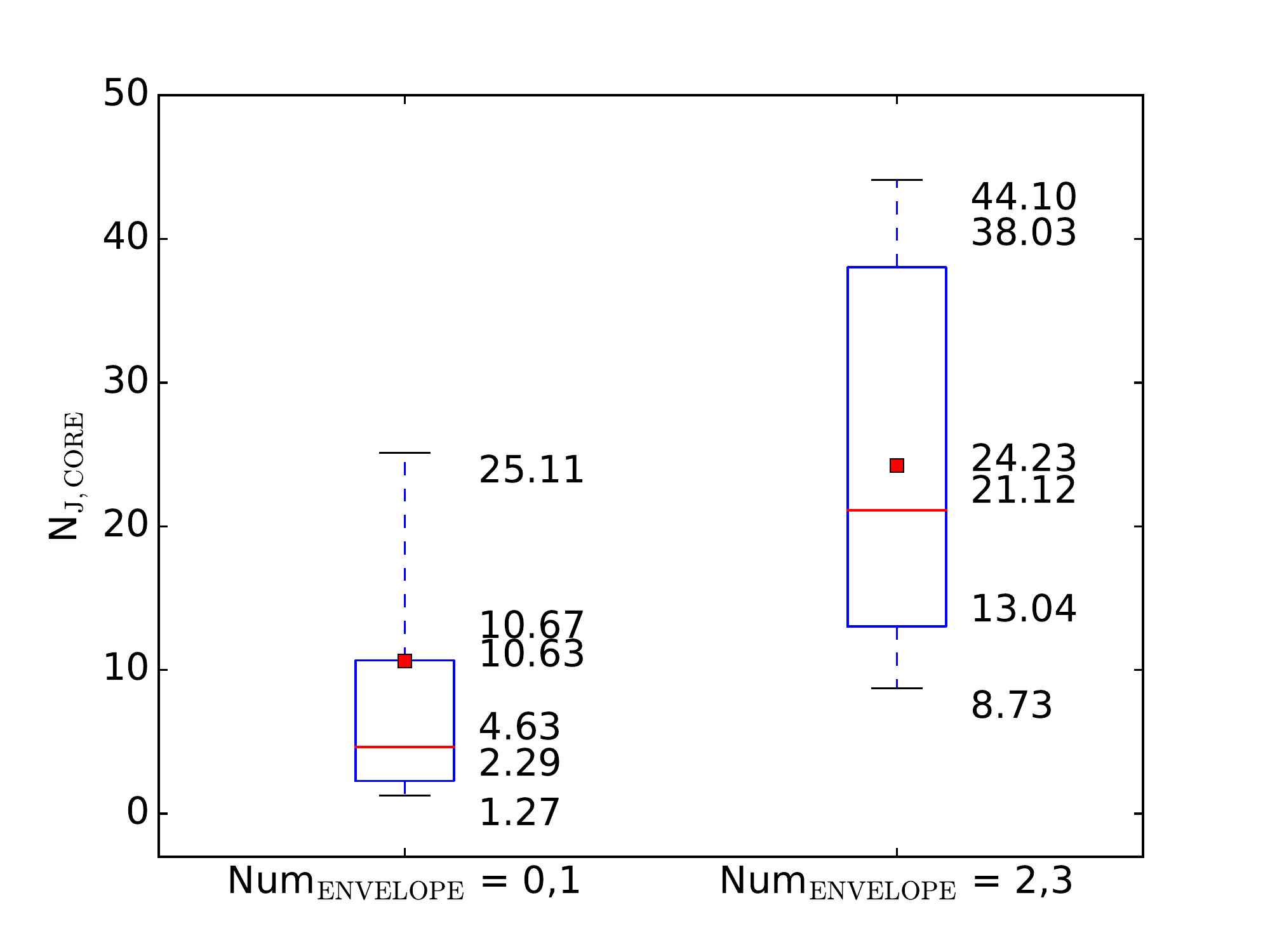}
\caption{Box and Whisker plot showing the distributions of the Jeans number of cores for two different population of enclosed envelopes. The first population constitutes the cores that have either 0 or 1 envelopes inside them. The second population constitutes the cores that have either 2 or 3 envelopes inside them. The numbers at the right side of the box and whisker diagram represent the 95$^{th}$ percentile, 3$^{rd}$ quartile, mean, median, 1$^{st}$ quartile and the 5$^{th}$ percentile going from the top to bottom respectively. Inside the box plot, the red square shows the value of mean and the red line shows the value of median.
\label{box_core}
}
\end{figure}

Figure \ref{box_core} represents the box and whisker plot for the distribution of the Jeans number of cores. The plot is shown for two different populations of cores. The first population consists of the cores that have either no envelopes or one envelope. The second population corresponds to cores with two or three envelopes. The p-value using K-S test in these two populations is $\sim$2 percent, so the distributions are significantly different within 95 percent confidence limit. Overall, Figure \ref{box_core} shows an increase in the number of enclosed envelopes with an increase in Jeans number of the cores.

\subsection{Envelope to protostellar objects} \label{envelopetodisk}

The envelope scale structures were probed with the SMA as part of the MASSES project. The protostellar objects were probed with the VLA as part of the VANDAM project. Below we explain the procedure in estimating mass and temperature of the SMA envelopes that are used to perform Jeans analysis in the envelopes.

\subsubsection{Envelope mass estimation} \label{mass estimation}

The mass of the SMA envelopes are estimated from the integrated flux of the SMA sources using Equation \ref{eq:2} \citep{Jorgensen07,Lee15} which converts 1.3 mm thermal dust emission into mass assuming that the emission is optically thin at 1.3 mm.

\begin{equation} \label{eq:2}
\begin{split}
M_{\rm{1.3~mm}} = 1.3~M_{\odot} \Bigg(\frac{F_{\rm{1.3~mm}}}{1~\rm{Jy}} \Bigg)\Bigg(\frac{D}{200~\rm{pc}} \Bigg)^2 \\
\times \Bigg \{\rm{exp} \Bigg[ 0.36 \Bigg( \frac{30~\rm{K}}{{\it T}_{\rm{d}}} \Bigg) \Bigg] - 1\Bigg\}
\end{split}
\end{equation}
where $F_{1.3 \rm{mm}}$ is the integrated flux density emitted by the source at 1.3 mm, $D$ is the distance to the source (230 pc) and $T_{\rm{d}}$ is the dust temperature of the envelopes. Equation \ref{eq:2} assumes the power law dust opacity which is calculated from  \cite{Ossenkopf94} with the models that have thin ice mantles coagulated at 10$^6$ cm$^{-3}$. We also assume the canonical gas-to-dust ratio of 100 \citep{Predehl95}.

To estimate $T_{\rm{d}}$, we used the model described in Equation 2 of \cite{Chandler00}. In brief, this model assumes a spherically symmetric envelope surrounding a central protostar and the temperature profile follows the power-law, $T~ \propto ~r^{-q}$ where $q$ is a function of dust emissivity ($\beta$), $q = 2/(4 + \beta)$, and $r$ is the distance of envelope from the central protostar. If $L_{\rm{bol}}$ is the bolometric luminosity of the protostar, the temperature of the envelope at a distance $r$ is,
\begin{equation}\label{eq:3}
T(r) = 60\Bigg(\frac{r}{2 \times 10^{15}~\rm{m}} \Bigg)^{-q} \Bigg(\frac{L_{\rm{bol}}}{10^5~L_{\odot}} \Bigg)^{q/2} \rm{K}
\end{equation}

Limited by the resolution of the SMA data, we calculated envelope temperature at a distance of 1000 AU from the central protostar. Similarly, consistent with the value of dust emissivity while calculating masses of the SMA sources, we used $q$ = 0.33. For $L_{\rm{bol}}$, we used the values from \cite{Tobin16}. Table \ref{envelopes_jeans} gives the temperature measurements at 1000 AU and the resulting masses for each envelope. The table also shows the group of envelopes with unreliable source fits. For these objects, the measured source properties such as mass, Jeans mass, etc are also unreliable. Such groups are designated as ``B" in Table \ref{envelopes_jeans}. In contrast, parameter estimates for the envelopes that belong to group ``A" are more robust. For further analysis below, we consider the envelopes that belong to group A only.

\subsubsection{Jeans Mass of Envelopes} \label{envelope jeans mass}

Table \ref{envelopes_jeans} gives the Jeans instability parameters for envelopes when they are supported by pure thermal motion and when they are supported by a combined thermal and non-thermal motion. For the pure thermal support, we used the mass and temperature estimates given in Table \ref{envelopes_jeans}. For the combined thermal and non-thermal support, we used N$_2$H$^+$ line width measurements from \cite{Kirk07}. The critical density of N$_2$H$^+$ is $\sim$10$^5$ cm$^{-3}$ and so it is suitable for studying the line width of envelopes in Perseus. We calculated the typical velocity dispersion at envelope scales as $\sim$0.13 km/s from line width measurements presented in Table 3 of \cite{Kirk07}.

\begin{figure}[tbh]
\centering
\vskip -1.0in
\includegraphics[scale=0.4]{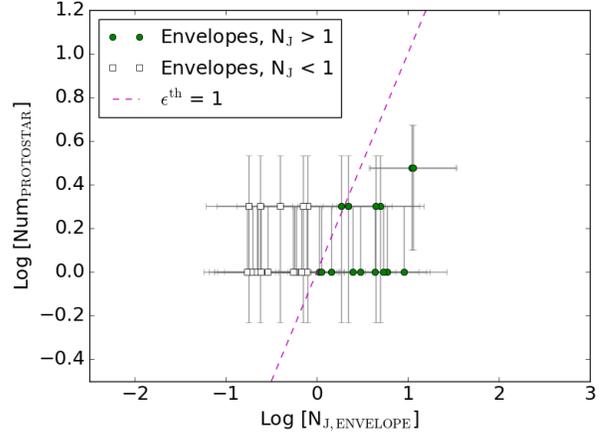}
\vskip -1.0in
\caption{Comparison of the number of protostellar objects with the Jeans number of the parent envelopes with thermal Jeans analysis. The green solid circles have N$_{\rm{J}}$ > 1 and the hollow squares have N$_{\rm{J}}$ < 1. The magenta dash line represents $\epsilon^{\rm{th}}$ = 1 line for perfect thermal Jeans fragmentation. Uncertainties in both axes are calculated similar to Figure \ref{NjVsN_cl}.
\label{NjVsN_envelope}
}
\end{figure}

Figure \ref{NjVsN_envelope} compares the number of VLA sources with the Jeans number of the SMA envelopes, assuming pure thermal support. The green solid circles in Figure \ref{NjVsN_envelope} represent the envelopes for which N$_{\rm{J}}$ $>$ 1 and the hollow square markers represent the envelopes for which N$_{\rm{J}}$ $<$ 1.
The median of the Jeans number of envelopes increase with the number of enclosed protostellar objects. Nevertheless, the robustness of this relation is limited by large uncertainties. There is a significant population of envelopes with N$_{\rm{J}}$ $<$ 1, which are less likely to fragment and form further stars. Hence, for further analysis we are only interested in the envelopes with N$_{\rm{J}}$ $>$ 1.

\startlongtable
\begin{deluxetable*}{lccccccccccc}
\centering
\tabletypesize{\scriptsize}
\tablecolumns{12}
\tablecaption{Jeans analysis in the envelopes \label{envelopes_jeans}}
\tablehead{\colhead{Envelope\tablenotemark{(a)}} & \colhead{Mass} & \colhead{Area ($\times 10^{-5}$)} & \colhead{T$_{1000AU}$} & \colhead{M$_{\rm{J}}$$^{\rm{th}}$\tablenotemark{(b)}} & \colhead{M$_{\rm{J}}$$^{\rm{th,nth}}$\tablenotemark{(b)}} & \colhead{N$_{\rm{J}}$$^{\rm{th}}$\tablenotemark{(c)}} & \colhead{N$_{\rm{J}}$$^{\rm{th,nth}}$\tablenotemark{(c)}} &  \colhead{Num$_{\rm{PROTOSTAR}}$} & \colhead{$\epsilon^{\rm{th}}$\tablenotemark{(d)}} & \colhead{$\epsilon^{\rm{th,nth}}\tablenotemark{(d)}$} & \colhead{Group\tablenotemark{(e)}}\\
\colhead{} & \colhead{[M$_{\odot}$]} & \colhead{[pc$^2$]} & \colhead{[K]} & \colhead{[M$_{\odot}$]} & \colhead{[M$_{\odot}$]} & \colhead{} & \colhead{} & \colhead{} & \colhead{} & \colhead{} & \colhead{}}
\startdata
B1-bN & 0.365 & 3.785798 & 17 & 0.08 & 0.11 & 4.3714 & 3.2674 & 1 & 0.23 & 0.31 & A \\
IC348 MMS1\tablenotemark{(f)} & 0.504 & 3.648606 & 22 & 0.1 & 0.12 & 4.9748 & 4.0686 & 2 & 0.4 & 0.49 & A \\
IC348 MMS2\tablenotemark{(f)} & 0.083 & 1.365433 & 22 & 0.12 & 0.15 & 0.6963 & 0.5695 & 1 & 1.44 & 1.76 & B \\
IRAS4B$'$ & 0.269 & 1.299492 & 26 & 0.08 & 0.09 & 3.3 & 2.8329 & 1 & 0.3 & 0.35 & B \\
L1448IRS3\tablenotemark{(g)} & 0.252 & 6.037167 & 29 & 0.32 & 0.36 & 0.7893 & 0.6998 & 1 & 2.53 & 2.86 & A \\
L1448NW\tablenotemark{(g)} & 0.163 & 3.69774 & 29 & 0.27 & 0.31 & 0.5953 & 0.5278 & 1 & 1.68 & 1.89 & A \\
L1451-MMS & 0.088 & 1.240456 & 12 & 0.05 & 0.07 & 1.9028 & 1.2451 & 1 & 0.53 & 0.8 & B \\
Per-bolo-45-SMM\tablenotemark{(h)} & 0.245 & 13.109837 & 12 & 0.16 & 0.25 & 1.5006 & 0.9819 & 0 & 0.0 & 0.0 & A \\
Per-bolo-58 & 0.213 & 10.513288 & 12 & 0.15 & 0.23 & 1.4362 & 0.9398 & 1 & 0.7 & 1.06 & A \\
Per-emb-1 & 0.337 & 3.011857 & 23 & 0.11 & 0.14 & 2.9991 & 2.4762 & 1 & 0.33 & 0.4 & A \\
Per-emb-2 & 0.897 & 1.390417 & 20 & 0.03 & 0.04 & 27.6627 & 22.0026 & 2 & 0.11 & 0.14 & B \\
Per-emb-3 & 0.079 & 2.858335 & 18 & 0.16 & 0.21 & 0.4888 & 0.3756 & 1 & 2.05 & 2.66 & B \\
Per-emb-5 & 0.357 & 2.248882 & 22 & 0.08 & 0.1 & 4.4256 & 3.592 & 2 & 0.68 & 0.84 & A \\
Per-emb-8 & 0.172 & 6.709789 & 24 & 0.31 & 0.37 & 0.5501 & 0.4626 & 1 & 1.82 & 2.16 & A \\
Per-emb-9 & 0.223 & 13.97958 & 19 & 0.33 & 0.43 & 0.6707 & 0.5211 & 1 & 1.49 & 1.92 & A \\
Per-emb-10 & 0.075 & 6.983227 & 19 & 0.34 & 0.44 & 0.2213 & 0.1719 & 1 & 4.52 & 5.82 & A \\
Per-emb-10-SMM & 0.024 & 1.399501 & 19 & 0.18 & 0.23 & 0.1364 & 0.106 & 0 & 0.0 & 0.0 & B \\
Per-emb-12 & 3.16 & 1.586092 & 29 & 0.03 & 0.04 & 99.6786 & 87.7338 & 2 & 0.02 & 0.02 & B \\
Per-emb-13 & 1.013 & 1.299492 & 26 & 0.04 & 0.05 & 24.1226 & 20.7087 & 1 & 0.04 & 0.05 & B \\
Per-emb-14 & 0.153 & 1.590404 & 19 & 0.08 & 0.1 & 1.8672 & 1.464 & 1 & 0.54 & 0.68 & B \\
Per-emb-15 & 0.097 & 4.269399 & 18 & 0.19 & 0.25 & 0.5156 & 0.3909 & 0 & 0.0 & 0.0 & A \\
Per-emb-16 & 0.131 & 6.885163 & 18 & 0.23 & 0.3 & 0.5668 & 0.4297 & 1 & 1.76 & 2.33 & A \\
Per-emb-17 & 0.1 & 5.881426 & 26 & 0.42 & 0.49 & 0.2367 & 0.2037 & 2 & 8.45 & 9.82 & A \\
Per-emb-18 & 0.202 & 6.523647 & 25 & 0.29 & 0.34 & 0.7003 & 0.5911 & 2 & 4.28 & 5.07 & A \\
Per-emb-19 & 0.021 & 1.362548 & 17 & 0.17 & 0.22 & 0.1262 & 0.095 & 1 & 7.93 & 10.52 & B \\
Per-emb-19-SMM\tablenotemark{(h)} & 0.005 & 1.362548 & 35 & 0.92 & 0.99 & 0.0059 & 0.0055 & 0 & 0.0 & 0.0 & B \\
Per-emb-20 & 0.058 & 3.666573 & 22 & 0.3 & 0.36 & 0.1947 & 0.1587 & 1 & 5.13 & 6.3 & A \\
Per-emb-20-SMM & 0.016 & 1.366165 & 22 & 0.27 & 0.33 & 0.0591 & 0.0482 & 0 & 0.0 & 0.0 & B \\
Per-emb-21 & 0.18 & 4.469851 & 25 & 0.23 & 0.27 & 0.7787 & 0.6573 & 1 & 1.28 & 1.52 & A \\
Per-emb-22 & 0.353 & 5.013858 & 26 & 0.19 & 0.22 & 1.8501 & 1.5805 & 2 & 1.08 & 1.27 & A \\
Per-emb-23 & 0.094 & 8.834424 & 20 & 0.39 & 0.49 & 0.2429 & 0.1919 & 1 & 4.12 & 5.21 & A \\
Per-emb-25 & 0.097 & 0.951497 & 21 & 0.08 & 0.1 & 1.2157 & 0.9825 & 1 & 0.82 & 1.02 & B \\
Per-emb-26 & 0.358 & 8.273704 & 30 & 0.34 & 0.38 & 1.0524 & 0.9335 & 1 & 0.95 & 1.07 & A \\
Per-emb-27 & 0.451 & 3.754182 & 34 & 0.2 & 0.22 & 2.2002 & 2.0156 & 2 & 0.91 & 0.99 & A \\
Per-emb-28 & 0.082 & 9.434481 & 18 & 0.37 & 0.49 & 0.223 & 0.1691 & 1 & 4.48 & 5.92 & A \\
Per-emb-29 & 0.41 & 4.606372 & 26 & 0.17 & 0.2 & 2.4561 & 2.1009 & 1 & 0.41 & 0.48 & A \\
Per-emb-30 & 0.052 & 0.673948 & 23 & 0.09 & 0.11 & 0.5724 & 0.4712 & 1 & 1.75 & 2.12 & B \\
Per-emb-33\tablenotemark{(g)} & 0.784 & 1.731601 & 29 & 0.07 & 0.08 & 11.0661 & 9.8107 & 3 & 0.27 & 0.31 & A \\
Per-emb-35 & 0.093 & 5.822 & 30 & 0.52 & 0.59 & 0.1786 & 0.159 & 2 & 11.2 & 12.58 & A \\
Per-emb-36 & 0.18 & 9.135532 & 27 & 0.46 & 0.53 & 0.3908 & 0.3398 & 2 & 5.12 & 5.89 & A \\
Per-emb-37 & 0.079 & 5.711896 & 18 & 0.27 & 0.36 & 0.2876 & 0.221 & 1 & 3.48 & 4.52 & A \\
Per-emb-40 & 0.027 & 1.597658 & 22 & 0.24 & 0.29 & 0.1125 & 0.092 & 2 & 17.78 & 21.74 & B \\
Per-emb-41 & 0.464 & 3.179652 & 19 & 0.08 & 0.1 & 5.8805 & 4.6105 & 1 & 0.17 & 0.22 & A \\
Per-emb-44 & 0.871 & 4.083969 & 21 & 0.08 & 0.09 & 11.4903 & 9.1933 & 3 & 0.26 & 0.33 & A \\
Per-emb-47 & 0.01 & 1.498126 & 21 & 0.35 & 0.43 & 0.0293 & 0.0237 & 1 & 34.07 & 42.16 & B \\
Per-emb-50 & 0.059 & 1.509439 & 35 & 0.3 & 0.33 & 0.196 & 0.1809 & 1 & 5.1 & 5.53 & B \\
Per-emb-51 & 0.24 & 2.081363 & 13 & 0.05 & 0.07 & 5.3183 & 3.5728 & 1 & 0.19 & 0.28 & A \\
Per-emb-53 & 0.062 & 3.384611 & 27 & 0.36 & 0.42 & 0.1717 & 0.1485 & 1 & 5.82 & 6.73 & A \\
Per-emb-54 & 0.128 & 6.045196 & 33 & 0.53 & 0.58 & 0.2412 & 0.2199 & 0 & 0.0 & 0.0 & A \\
Per-emb-56 & 0.018 & 1.137566 & 19 & 0.17 & 0.22 & 0.1073 & 0.0828 & 1 & 9.32 & 12.07 & B \\
Per-emb-57 & 0.046 & 1.517082 & 14 & 0.09 & 0.13 & 0.5252 & 0.3596 & 1 & 1.9 & 2.78 & B \\
Per-emb-58\tablenotemark{(h)} & 0.01 & 1.489637 & 19 & 0.3 & 0.38 & 0.0327 & 0.0255 & 1 & 30.58 & 39.24 & B \\
Per-emb-61 & 0.018 & 1.440329 & 16 & 0.17 & 0.23 & 0.1079 & 0.0792 & 0 & 0.0 & 0.0 & B \\
Per-emb-62 & 0.077 & 1.473029 & 23 & 0.14 & 0.17 & 0.5601 & 0.4624 & 1 & 1.79 & 2.16 & B \\
Per-emb-63 & 0.018 & 1.557983 & 23 & 0.3 & 0.36 & 0.061 & 0.0505 & 1 & 16.4 & 19.8 & B \\
Per-emb-64 & 0.041 & 1.592964 & 25 & 0.23 & 0.27 & 0.1798 & 0.1527 & 1 & 5.56 & 6.55 & B \\
Per-emb-65 & 0.047 & 1.489332 & 15 & 0.1 & 0.14 & 0.485 & 0.3461 & 1 & 2.06 & 2.89 & B \\
SVS13B & 0.889 & 5.917212 & 21 & 0.1 & 0.12 & 8.9669 & 7.1743 & 1 & 0.11 & 0.14 & A \\
SVS13C & 0.199 & 4.127341 & 22 & 0.18 & 0.22 & 1.1279 & 0.9224 & 1 & 0.89 & 1.08 & A \\
\enddata
\tablenotetext{(a)}{ Envelopes are the SMA sources and their nomenclature is adopted from \cite{Tobin16} for consistency. We could not find some sources in the literature so we designated them "SMM" at the end of their name. For example, Per-bolo-45-SMM is a new detection that does not lie in the same region as Per-bolo-45. All the envelopes are detected at 5-$\sigma$ contour, unless otherwise stated.}
\tablenotetext{(b)}{ Jeans mass considering thermal (M$_{\rm{J}}$$^{\rm{th}}$) and total support (M$_{\rm{J}}$$^{\rm{th,nth}}$).}
\tablenotetext{(c)}{ Jeans number considering thermal (N$_{\rm{J}}$$^{\rm{th}}$) and total support (N$_{\rm{J}}$$^{\rm{th,nth}}$).}
\tablenotetext{(d)}{ Efficiency is calculated by taking ratio of the number of protostellar objects to the Jeans number of envelopes considering both thermal ($\epsilon^{\rm{th}}$) and combined ($\epsilon^{\rm{th,nth}}$) support.}
\tablenotetext{(e)}{ Refer Table \ref{envelopes}. A: Reliable fits, B: Unreliable fits.}
\tablenotetext{(f)}{ Nomenclature adopted from \cite{Lee16}.}
\tablenotetext{(g)}{ SMA envelope is detected at 6$\sigma$ contour.}
\tablenotetext{(h)}{ SMA envelope is detected at 4$\sigma$ contour.}
\end{deluxetable*}

\begin{figure}[tbh]
\centering
\vskip -1.1in
\includegraphics[scale=0.4]{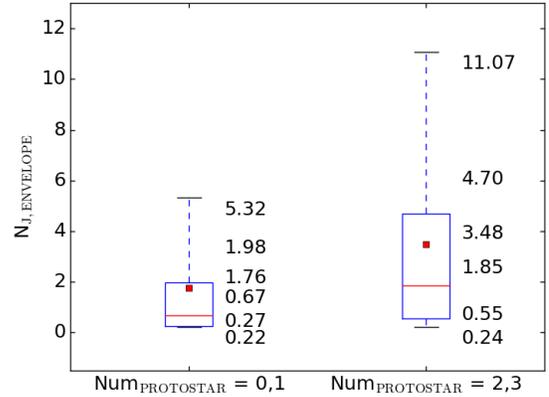}
\vskip -1.0in
\caption{Box and Whisker plot showing the distribution of the Jeans number of envelopes for two different population of enclosed disk scale objects. The first population constitutes the envelopes that have either 0 or 1 disk scale objects inside them. The second population constitutes the cores that have either 2 or 3 disk scale objects inside them. The numbers at the right side of the box and whisker diagram represent the 95$^{th}$ percentile, 3$^{rd}$ quartile, mean, median, 1$^{st}$ quartile and the 5$^{th}$ percentile going from the top to bottom respectively. Inside the box plot, the red square shows the value of mean and the red line shows the value of median. The p-value obtained from the K-S test is $\sim$50\%, implying that these two populations are not significantly different.
\label{box_envelope}
}
\end{figure}

Figure \ref{box_envelope} shows the distribution of the Jeans number of envelopes that belong to group "A" in a box and Whisker plot. For this plot, we have two populations of envelopes, with the first population have 0 or 1 protostellar objects while the other population has 2 or 3 protostellar objects. The median values are $\sim$0.67 for the first population and $\sim$1.85 for the second. The representation of statistics at the right of the box diagram is same as Figure \ref{box_core}. The p-value obtained from the K-S test for these two populations is  $\sim$50 percent. Thus, unlike the previous hierarchy, we cannot statistically distinguish between the distributions of the Jeans numbers for the two envelope populations.

\section{Combining all hierarchies} \label{combination}

\begin{figure}[tbh]
\centering
\vskip -1.0in
\includegraphics[scale=0.4]{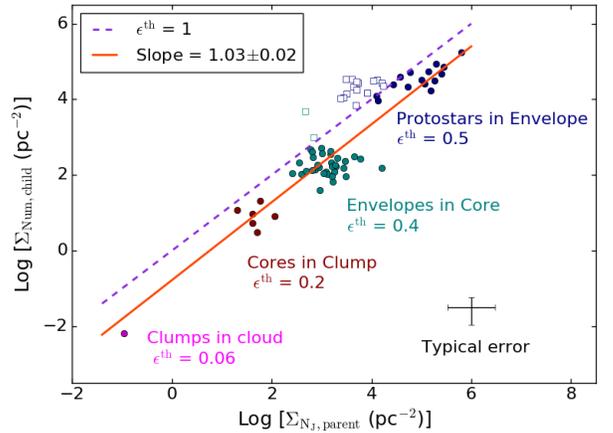}
\vskip -1.0in
\caption{Surface density plot combining all the hierarchies. The x-axis shows the Jeans number surface density of the parent structure  (Jeans number of parent/ area of parent) and the y-axis shows the number surface density of the child structure (number of children objects / area of parent). The data shown in different colors represent clumps in cloud, cores in clump, envelopes in core, and the protostellar objects in envelopes. The solid circles have $N_{\rm{J}}$ > 1 and hollow squares have $N_{\rm{J}}$ < 1. The dash line represents a $\epsilon^{\rm{th}} = 1$ line, where the number of children objects are equal to the Jeans number of parent objects. Solid line shows the linear best fit for all the data for cloud, clump and cores, and for $N_{\rm{J}}$ > 1 data for envelopes. Slope of the best fit line is $\sim$1.
\label{surfden_th}
}
\end{figure}

We examined the hierarchical structure in Perseus from cloud scales to protostellar objects in \S \ref{jeansanalysis}.  In general, we find a correlation between Jeans number and the number of children objects, where parent structures with higher Jeans numbers have more substructure.  To illustrate the multiscale correlation, Figure \ref{surfden_th} combines the results in each hierarchy in a single plot. Figure \ref{surfden_th} compares the Jeans number of each parent structure with their number of children objects, with both values shown as a surface density.
If we plot the number of child objects with the Jeans number of parent objects for all the scales without dividing by area, the data overlap with each other because of the small range of Jeans number of parent objects and the number of child objects (see Figures \ref{NjVsN_cl}, \ref{NjVsN_core}, and \ref{NjVsN_envelope}). In such a plot, different physical scales from cloud to protostellar objects cannot be visualized, which motivated the need to separate them by dividing by the area. Since the five scales of hierarchy vary widely in terms of their physical scale, we used a surface density plot to visualize each scale distinctly.

In Figure \ref{surfden_th}, the solid circles represent structures with $N_{\rm{J}}$ $>$ 1, and the hollow squares show the data for which $N_{\rm{J}}$ $<$ 1 for the parent population. The typical uncertainty is shown in lower right region of Figure \ref{surfden_th}. The dash line shows the $\epsilon^{\rm{th}} = 1$ relation for perfect thermal fragmentation. Solid line represents the best fit line for all the data for the scales of cloud, clump and core, and $N_{\rm{J}}$ $>$ 1 data for the envelopes as noted in \S \ref{envelope jeans mass}. The best fit results do not change if we include $N_{\rm{J}}$ $>$ 1 criteria for fitting at all scales as there are only two other cores that have $N_{\rm{J}}$ $<$ 1. We used the Markov Chain Monte Carlo (MCMC) method \citep{van03} to fit a linear model to the data. MCMC uses random numbers from a Markov Chain to characterize a probability distribution. We fit the relation within the uncertainties for the different scales. For the details of the use of MCMC in fitting astronomical data, see \cite{pokhrel16}. For the underlying assumption and choice of priors, we followed \cite{pokhrel16} and we used the PYTHON package $pymc$ \citep{patil10} to apply the MCMC method.

Figure \ref{surfden_th} assumes that the thermal gas motions are solely responsible for stability of the structure against gravitational collapse. The slope of the best fit line is 1.03 $\pm$ 0.02, with a Pearson correlation coefficient of 0.95. The best fit line is close to but offset from the $\epsilon^{\rm{th}} = 1$ line relation which implies that only a fraction ($<$ 1) of the mass in the parent structure has been converted into children structures. This lower efficiency is similar to the result seen for individual hierarchy levels in Figures \ref{NjVsN_cl}, \ref{NjVsN_core} and \ref{NjVsN_envelope}, and is similar to the result of \cite{Palau15} for cores.

We estimated the formation efficiency of children objects for each level of hierarchy ($\epsilon^{\rm{th}}$) using the process described in \S \ref{jeansanalysis}. We found average values of $\epsilon^{\rm{th}}$ as 0.06, 0.2, 0.4 and 0.5 for the formation of clumps, cores, envelopes and protostars respectively (see Figure \ref{surfden_th}), but these scales can also have a broad range.  For example, we found CFEs between 0.04 and 0.6 assuming their parent clumps are thermally supported (see \S \ref{clumptocore}) which is similar to the value of CFE from other independent measurements \citep{Bontemps10,Palau13,Palau15}. If we exclude the two cores for which $N_{\rm{J}}$ $<$ 1, the $\epsilon^{\rm{th}}$ for cores is $\sim$0.2, however the power-law relation stays the same. Thus, we find that thermal support alone cannot predict the amount of fragmentation detected on cloud or clump scales, while there is better agreement on the scales of cores and envelopes. Nonetheless, the tendencies for $\epsilon^{\rm{th}} < 1$ and for $\epsilon^{\rm{th}}$ to increase with decreasing size scale remain to be explained.

Figure \ref{surfden_th} shows an increasing trend in thermal efficiencies towards smaller scales. To test the robustness of this trend, we calculated the uncertainty in typical thermal efficiency with a Monte Carlo approach. Since efficiency is the ratio of the number of children objects to the Jeans number of the parent object, the uncertainty in the efficiency is dominated by the uncertainty in the Jeans number. The Jeans number is certain within a factor of $\sim$3, whereas the number count of children objects follow Poisson statistics as an upper limit uncertainty. Thus the efficiencies are varied randomly within a factor of 3-4 to simulate a range of datasets within the errors. We used 5000 iterations for each level and for each iteration we calculated the average efficiency. Finally we computed the standard deviation of all the simulated average efficiencies to find the uncertainties at each scale. The thermal efficiencies have uncertainties of 0.05, 0.08, 0.16 and 0.11 for the clumps, cores, envelopes and protostars respectively.

\begin{figure}[tbh]
\centering
\vskip -1.0in
\includegraphics[scale=0.4]{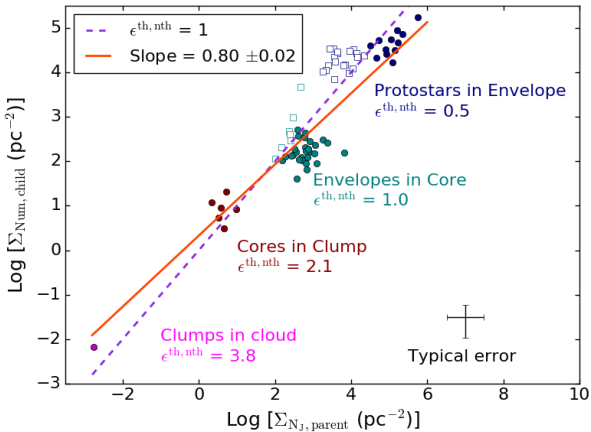}
\vskip -1.0in
\caption{Surface density plot similar to Figure \ref{surfden_th}, now considering the combined support of both the thermal and non-thermal motions against gravitational collapse.
\label{surfden_tot}
}
\end{figure}

As a companion to Figure \ref{surfden_th}, Figure \ref{surfden_tot} assumes that the means of support for cloud stability is the combination of both thermal and non-thermal motion of gas. We performed Jeans analysis for the combined support using the process described in \S \ref{jeansanalysis}. The best fit is performed on the structures with $N_{\rm{J}}$ $>$ 1. For the combined support, the formation efficiency $\epsilon^{\rm{th,nth}}$ decreases while going from large to small scales. We found $\epsilon^{\rm{th,nth}}$ as 3.8 $\pm$ 2.9, 2.1 $\pm$ 0.8, 1.0 $\pm$ 0.4 \& 0.5 $\pm$ 0.1 for the formation of clumps, cores, envelopes and protostellar objects. The uncertainties in $\epsilon^{\rm{th,nth}}$ are obtained using Monte Carlo method similar to the pure thermal case. Thus the combined thermal and non-thermal support follow a different (and opposing) trend as the thermal only case.

It is interesting to note that for the formation of protostellar objects inside envelopes, the case with combined thermal and non-thermal support gives a very similar efficiency as the thermal only case. This implies that the non-thermal motions are relatively insignificant at these scales and fragmentation is entirely driven by the competition between gravity and thermal support. As we move towards the larger scales, the combined efficiency is greater than unity and hence unphysical.

Finally, we performed best fit in Figures \ref{surfden_th} and \ref{surfden_tot} by including all the data. This includes that data with $N_{\rm{J}}$ $>$ 1 and also $N_{\rm{J}}$ $<$ 1. The hierarchy level concerning envelope to protostellar objects has the most data with $N_{\rm{J}}$ $<$ 1. Thus the values of $\epsilon^{\rm{th}}$ and $\epsilon^{\rm{th,nth}}$ are changed for the envelope scale. $\epsilon^{\rm{th}}$ changes from 0.5 (with $N_{\rm{J}}$ $>$ 1 data only) to 2.1 (including all data) and $\epsilon^{\rm{th,nth}}$ changes from 0.5 to 2.6. These values are similar within their uncertainties. However the results with efficiencies greater than unity are unrealistic to explain.

\begin{figure}[tbh]
\centering
\includegraphics[scale=0.4]{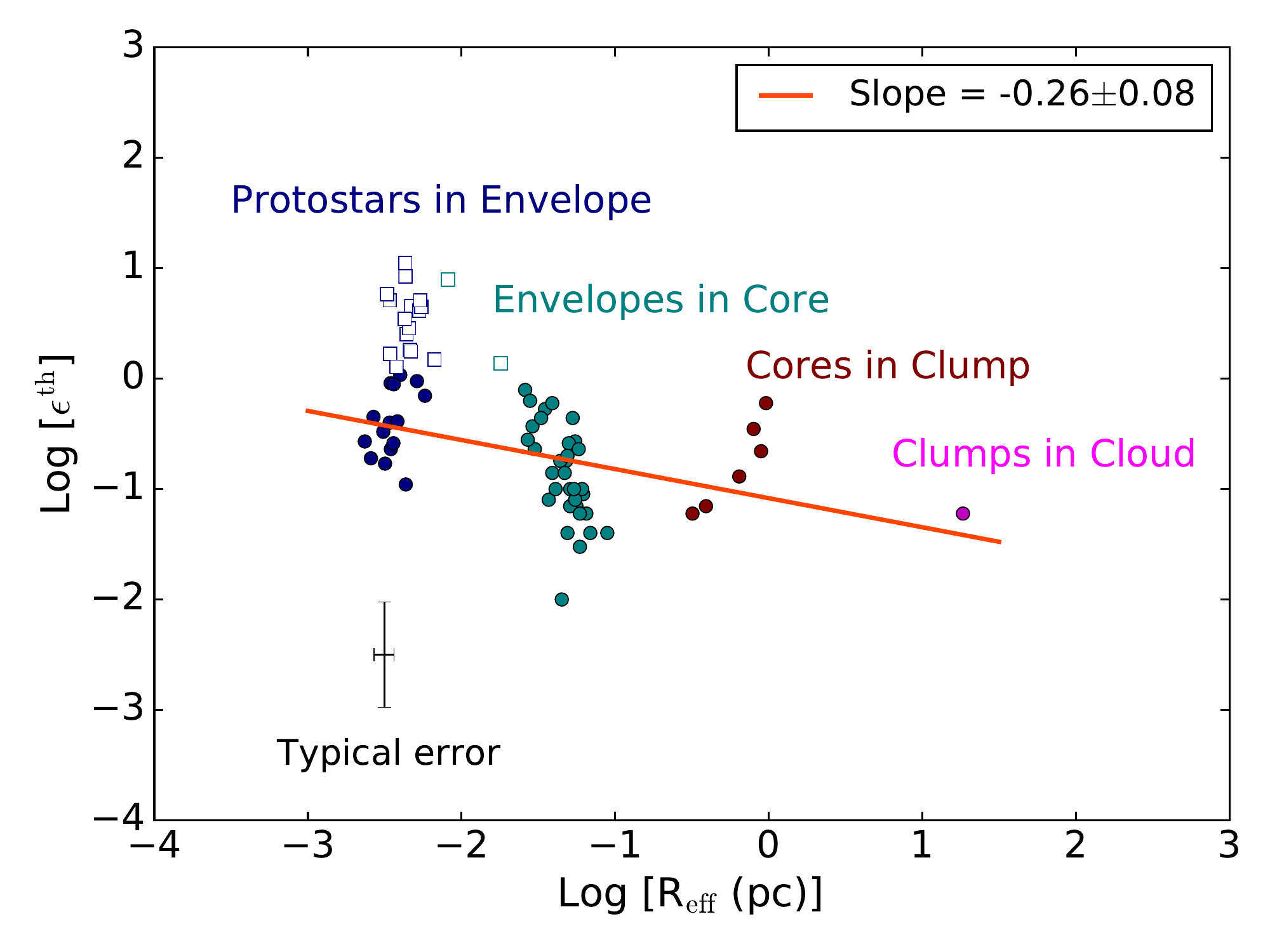}
\caption{Comparison of the thermal efficiency with the size of parent structures. The size is represented in terms of an effective radius assuming a spherical geometry. Symbols are the same as in Figure \ref{surfden_th}. The typical error bar on the data is shown on lower left side of the plot.
\label{effVsReff}
}
\end{figure}

The good correlation in Figure \ref{surfden_th} and \ref{surfden_tot} is due largely to the fact that the range of $\epsilon^{\rm{th}}$ is much smaller than the range of area or surface density. We stress that the point of making these surface density plots is not to claim any kind of correlation between the Jeans number of parent objects and number of child objects. Rather, we include these plots only to show that there is sub-thermal efficiency at each scale. The dependence of $\epsilon^{\rm{th}}$ on size scale, without normaliation by area, is shown in Figure \ref{effVsReff}.

To remove the possible degeneracy introduced by area in the surface density plot, in Figure \ref{effVsReff} we plotted the thermal efficiency of each parent object with their effective radius. For cloud, clump and core scale we calculated the effective radius by assuming spherical geometry of the structures. For envelope scales the effective radius is the geometric mean of major and minor axes of the source. The solid line in Figure \ref{effVsReff} represents the best fit line with slope of a power-law -0.26 $\pm$ 0.08. The data for best fit is taken to be same as in Figure \ref{surfden_th}. If we fit the data with $N_{\rm{J}}$ $>$ 1 for all the scales, the slope is -0.23 $\pm$ 0.07 which is within the uncertainty range of -0.26 $\pm$ 0.08. The plot explicitly depicts the increasing trend of thermal efficiency value for smaller objects. The $\epsilon^{\rm{th}}$ is maximum for protostars and gradually decreases when we probe larger scales. Thus we can conclude that as the size scale decreases in structures in a molecular cloud, the efficiency of thermal fragmentation increases.

\section{Discussion} \label{conclusions}

Our study shows that fragmentation is a scale dependent process. The mass of the structures in upper level hierarchy such as cloud and clumps are higher than the thermal Jeans mass. In contrast, masses are around thermal Jeans mass for the envelope and disk scales which provides a further support to the idea of thermal fragmentation at smaller scales. This provides further clue that we may be reaching the coherence level while going from the cores to envelope when the role of thermal fragmentation starts dominating. 

In the later stage the fragmentation process in low mass star forming regions seem to be controlled mostly by the gravitational contraction with the decrease of the thermal Jeans mass with an increase in density during contraction. However to conclude this statement we need to analyze the magnetic field contribution as well in the future. Nevertheless, our work supports the view that the thermal motion can provide support against gravity and stabilize the cloud sub-structure, especially at smaller scales.

Our results are consistent with some recent studies for cores \citep{Miettinen12, Palau15,Busquet16} that supports the notion of thermal Jeans fragmentation over non-thermal fragmentation. \cite{Miettinen12} detected low mass class 0 protostellar fragments inside the SMM6 core in B9 region of the orion molecular cloud and conclude that the origin of the substructure is due to thermal Jeans fragmentation. Similarly, \cite{Palau15} studied 19 dense cores in nearby molecular clouds and found that most of the fragments detected in their sample are around the thermal Jeans limit. A more recent study by \cite{Palau17} in the Orion Molecular Cloud 1 South (OMS-1S) shows that fragmentation from 100 AU to 40 AU is also consistent with thermal Jeans processes. Thus, Jeans fragmentation seems to be a viable process in some high mass star forming regions as well (e.g., \citealt{Samal15}).

On the other hand, our results do not appear to agree with some studies of higher mass IRDCs (\citealt{Zhang09,Pillai11,Wang14,Lu15}). They find that the fragments have masses much larger than the thermal Jeans mass and are consistent with the non-thermal Jeans mass. However, this is similar to our results in massive clumps. Hence thermal fragmentation may be dominant only in low-intermediate star forming regions. This suggests that although non-thermal motion seems important for fragmentation and the formation of massive cores in a cluster, the low mass cores may be produced by thermal fragmentation. Indeed, \cite{Zhang15} reported a population of low mass cores in a protocluster using more sensitive observations with ALMA, which appears to be consistent with thermal fragmentation. In another study, \cite{Lu15} find cores more massive than Perseus cores in clumps with $\epsilon^{\rm{th}}$ = 0.01 - 0.02 and $\epsilon^{\rm{th,nth}}$ = 0.2 - 0.3. In contrast, in their simulation work \cite{offner16} reported that fragmentation was less common in lower mass cores, where thermal pressure was more important (relative to turbulence and magnetic pressure). It is important to compare Jeans fragmentation in high and low mass clouds in more detail.

We performed Jeans analysis at all the scales in the hierarchy, comparing the Jeans number of parent object with number of child objects.  An alternative procedure would be to compute the effective  critical Bonnor-Ebert (BE) mass \citep{Ebert55,Bonnor56} for each parent structure.  This mass has the same dependence on temperature and density as the Jeans mass, but its value is less by a factor 2.47 \citep{Mckee07}. We calculated the BE efficiencies considering BE mass and we found $\epsilon^{\rm{th}}$ as 0.02 $\pm$ 0.02, 0.08 $\pm$ 0.03, 0.16 $\pm$ 0.06 and 0.35 $\pm$ 0.08 from the cloud scale to protostellar objects, which are within the uncertainty limit of $\epsilon^{\rm{th}}$ that we obtained with Jeans analysis. Moreover, using the BE mass would be less convenient for comparing results with many previous studies which rely on the Jeans mass (for example \citealt{Zhang09,Pillai11,Miettinen12,Lu15,Palau15,Busquet16,Palau17}). Therefore in this paper we use the Jeans mass rather than the BE mass.

Our use of non-thermal velocity dispersion derived from line widths provides a comparison between the Jeans number based on this velocity dispersion, the Jeans number based on the gas kinetic temperature, and the number of observed fragments. This comparison is a test of which velocity dispersion gives better agreement with observed fragment numbers, but it is not a test of fragmentation in MHD turbulence-regulated star formation models and simulations (for example by \citealt{Padoan99,Padoan02,MacLow04,Hennebelle11}, etc). Such numerical models represent motions which are more anisotropic, time-varying, magnetized, and scale-dependent than those analyzed here with simple models of Jeans fragmentation.

Similar observational works on hierarchical fragmentation in other nearby molecular clouds are needed to further test our results. These works should be further extended to the massive star forming regions where the relative importance of non-thermal motions may be different from Perseus. Also, a detailed comparison with simulations of low mass star forming regions is necessary to further constrain the role of thermal and non-thermal support in both the smaller scales such as protostars and disks, and the larger scale such as the cloud and clumps.

\section{Conclusion} \label{summary}

In this study, we examined the multiscale structure in the Perseus molecular cloud from the scale of the cloud ($\geq$ 10pc) to the scale of dust and ionized gas around protostars ($\sim$15 AU). To study the scales of the cloud, clump, core and disk scale objects, the data is derived from the available literature, and for the scale of envelopes we used new SMA data from the MASSES project. This breadth of scale is unique to this study and reveals how clouds themselves are structured from large to small scales.

We traced 5 distinct scales and compared the number of fragments seen in each child structure with the expected number that could be produced by the parent structure according to Jeans fragmentation. We first considered purely thermal Jeans fragmentation. For such system we found a positive correlation between the number of children objects and the Jeans number of their parent objects at all scales. This trend, however, is not one-to-one. The average number of children objects are always less than the Jeans number of parent object. Under pure thermal support, the efficiency of the structure formation is 0.06, 0.2, 0.4 and 0.5 for clumps, cores, envelopes and protostellar objects. Thus thermal motions are least efficient in providing support at larger scales such as the whole cloud, and most efficient at smaller scales such as the protostellar objects.

Considering the combined support of both thermal and non-thermal motions, the efficiency of formation is largest and unphysical ($>$1) for the clumps, cores and envelopes, and least for the protostellar objects. We quantified the combined efficiency as 3.8, 2.1, 1.0 and 0.5 for the formation of clumps, cores, envelopes and protostellar objects. For the protostellar objects, both $\epsilon^{\rm{th,nth}}$ and $\epsilon^{\rm{th}}$ have value $\sim$0.5, which shows that the thermal support is significant at these scales, however this doesn't rule out the possibility of other means of support such as magnetic pressure.

\acknowledgments
R.P. acknowledges the support from the NASA Grant NNX14AG96G. J.J.T. acknowledges support from the Netherlands Organization for Scientific Research (NWO) from Veni grant 639.041.439. We thank the anonymous referee for their comments and suggestions. We also thank Seyma Mercimek for fruitful discussion regarding core distribution in the clumps. SMA is a joint project between the Smithsonian Astrophysical Observatory and the Academia Sinica Institute of Astronomy and Astrophysics and funded by the Smithsonian Institution and the Academia Sinica. The authors thank the SMA staff for executing these observations as part of the queue schedule, Charlie Qi and Mark Gurwell for their technical assistance with the SMA data, and Eric Keto for his guidance with SMA large-scale projects. 

\software{We have used the following software packages for this study- 
		APLpy \citep{Robitaille12},
        AstroPy \citep{astropyref},
		Matplotlib \citep{Hunter07},
		MIR \url{(https://www.cfa.harvard.edu/~cqi/mircook.html)}, 
		MIRIAD \citep{Sault95},
        NumPy \url{(https://doi.org/10.1109/MCSE.2011.37)},
        pymc \citep{patil10}, 
        SciPy \citep{scipyref}.
        }


\bibliographystyle{yahapj}
\bibliography{references}


\end{document}